\documentclass[12pt]{article}

\usepackage{a4,graphics,graphicx,amsmath,amssymb,cite}
\usepackage[verbose]{wrapfig}

\catcode`@=11 \@addtoreset{equation}{section} \catcode`@=12

\begin{document}
\title{
{\baselineskip -.2in
\vbox{\small\hskip 4in \hbox{IITM/PH/TH/2010/10}}} 
\vskip .4in
\vbox{
{\bf \large Instanton Corrected Non-Supersymmetric Attractors }
}
\author{Pramod Dominic${}$\thanks{email: pramod@physics.iitm.ac.in}
~and
Prasanta K. Tripathy${}$\thanks{email: prasanta@physics.iitm.ac.in} \\
\normalsize{\it ${}$ Department of Physics,}\\
\normalsize{\it Indian Institute of Technology Madras} \\
\normalsize{\it  Chennai 600 036, India.} 
}}
\maketitle
\begin{abstract}
We  discuss non-supersymmetric attractors with an instanton correction in Type $II A$ string theory compactified on a Calabi-Yau three-fold at large volume. For a stable non-supersymmetric black hole, the attractor point must minimize the effective black hole potential. We study the supersymmetric as well as non-supersymmetric attractors for the D0-D4 system with instanton corrections. We  show that in simple models, like the STU model,  the flat directions of the mass matrix can be lifted by a suitable choice of the instanton parameters. 
\end{abstract}
\renewcommand{\thefootnote}{\arabic{footnote}}
\setcounter{footnote}{0}
\newpage
\section{Introduction}
 
   Some time back, in a seminal work \cite{Ferrara:1995ih}, Ferrara, Kallosh and Strominger 
observed a remarkable feature of asymptotically flat, static, spherically symmetric, extremal black 
holes in $N=2$ supergravity coupled to $n$ vector multiplets in four dimensions. Analyzing the 
spinor conditions, these authors found that the black hole solutions behave as attractors. 
Although, the scalar fields arising from the vector multiplets take arbitrary values at spatial 
infinity, they run into a fixed point at the horizon of the black hole. The values of the scalar 
fields at the fixed point are determined by the magnetic charges of the black hole. This result
has subsequently been generalized for dyonic black holes by Strominger  \cite{Strominger:1996kf}.

It was soon realized that this apparently simple looking result has remarkable consequences
in several aspects of supersymmetric black holes. For example, one might naively 
think that since the black hole solution in supergravity in general involves scalar fields apart from 
gauge fields, the area of the black hole and hence it's entropy should depend on these fields.
In particular, the macroscopic entropy should depend on the boundary condition imposed on the 
scalar fields at infinity. However,  a microscopic counting of black  hole entropy indicates that 
it depends only upon the electric and magnetic charges of the black hole which are quantized 
objects. Since the scalar field vacuum expectation values are continuous variables this is 
a seemingly paradox. This paradox is resolved by the attractor mechanism. 

Various properties related to the attractor mechanism in extremal black holes have  
been investigated extensively afterwards \cite{Ferrara:1996dd,Ferrara:1996um,Gibbons:1996af,Ferrara:1997tw}. It has been shown that the central charge of the supergravity 
theory is extremised 
at the attractor point and the macroscopic entropy of the black holes can be derived from the 
value of the central charge at the extremum. The problem of finding extremal black hole has 
been mapped on to an effective one-dimensional theory and the attractor is described in terms
of the extremum of the potential in the effective theory. Similar results in $N=4$ and $N=8$ 
supergravity have also been established in some of the above references.  Some of the early results on attractor equations in the presence of higher derivative terms are discussed in 
\cite{Behrndt:1998eq, LopesCardoso:1998wt,LopesCardoso:1999cv,LopesCardoso:1999ur}. 
These results played a key role in arriving at the OSV conjecture on the connection between 
the macroscopic entropy  of black holes and the topological partition function \cite{Ooguri:2004zv}. 
Multi-centered 
configurations have been studied  and the connection between BPS states in string theory has 
also been discussed \cite{Denefa,Denef:2001xn}.

Another area where important development has taken place in recent times is in the study 
of the  attractor mechanism for non-supersymmetric extremal black holes. It has been realized 
that the attractor mechanism is actually a consequence of extremality of the black hole and not 
supersymmetry! Although the existence of a 
critical point in the moduli space, which need not preserve supersymmetry, had already been
pointed out long back \cite{Gibbons:1996af,Ferrara:1997tw}, not much attention was paid to this 
class of configurations. A thorough investigation of  the attractor mechanism in non-supersymmetric
black holes was done only as late as  2005 by Goldstein {\it et.al.}, \cite{gijt}.  The existence of non-supersymmetric attractors in string theory was first shown in \cite{pkt1}. Subsequently, the attractor 
equations were written down more elegantly in terms of algebraic equations and their similarity 
with flux vacua has also been established \cite{Kallosh:2005ax,Kallosh:2006bt}.

Although the non-supersymmetric extremal black holes studied in \cite{pkt1} shared many 
features with their supersymmetric counterparts, there is one striking difference. The black holes 
preserving supersymmetry are stable. However, for the non-supersymmetric solution, it turns 
out that, out of the $n$ complex scalar fields only $(n+1)$ real scalar fields are massive and the
remaining $(n-1)$ real scalars have vanishing quadratic terms at the attractor point. The cubic 
terms are absent. And hence the attractors become stable or unstable depending on whether 
the quartic terms become positive-definite or not \cite{Nampuri:2007gv}. In some cases, 
as for instance in the STU model, the potential in these $(n-1)$ real scalars remain exactly flat. 

The appearance of these massless fields is because only leading order terms in the string theory 
are considered. In the present work we explore the possibility of lifting these flat directions
 by including quantum corrections. In some of the simple cases, the exact prepotential for string 
 theory on various one and two-parameter families of Calabi-Yau models, (including the instanton
 effects), has been computed by using mirror symmetry 
 \cite{Suranyi:1992kf,Font:1992uk,Klemm:1992tx,Candelas:1993dm,Candelas:1994hw,Govindarajan:2000vi}.  
 We examine one of the simplest two-parameter Calabi-Yau 
 manifolds considered in these papers and study the behavior of non-supersymmetric attractors
 arising from $IIA$ compactification on it. We then generalize the analysis to compactification
 on arbitrary Calabi-Yau manifolds, but for simplicity, we retain only the instanton contributions.
 We show that the instanton contribution to the prepotential lifts these massless directions.

 This paper is organized as follows: In \S2 we give an introduction to non-supersymmetric attractors. First we discuss non-supersymmetric attractors in the context of $N=2$ theories. 
  We summarize the basic formalism  in terms of the effective potential, including the conditions associated with the non-supersymmetric attractors. 
 There we outline the construction of the effective potential in terms of the super-potential and K\"ahler potential and the computation of the black hole entropy in  terms of the scalars and charges of the black hole. We then consider non-supersymmetric attractors in type ${II} A$ string theory compactified on a Calabi-Yau three-fold and  we discuss the computation of supersymmetric
  and non-supersymmetric black hole solution from the from the effective potential. We compute the mass matrix and its spectrum for the  D0-D4 system.

 In \S3 we discuss a simple model of  non-supersymmetric attractors for type $IIA$ string theory compactified in a simple two-parameter Calabi-Yau manifold. In Section 4 we consider non-supersymmetric attractors with an instanton correction in type ${II} A$ string theory compactified on an arbitrary Calabi-Yau three-fold. Here we construct the effective potential in the context of instantons. We obtain both  the supersymmetric as well as the non-supersymmetric attractor solutions for the extremal black hole. We also  construct the mass matrix by computing the second derivatives of the effective potential. 
We find the spectrum of the mass  matrix in simple models such as the STU model. We show that with a suitable choice of instanton parameters we will can lift the flat directions of the mass matrix using numerical analysis and degenerate perturbation theory. Finally we conclude by summarizing the main results of the paper followed by an outline of future work.

  \section{Non-supersymmetric attractors}

In this section we will give a brief introduction to attractor mechanism in four dimensional $N=2$ 
supergravity  theory arising from compactification of string theory on Calabi-Yau three-folds. The 
supersymmetric attractors were first studied by Ferrara et al  \cite{Ferrara:1995ih} by solving the 
spinor conditions.  Subsequently the Euler-Lagrange equations were analyzed in detail and an 
effective one-dimensional description were given  \cite{Ferrara:1997tw}. A detail analysis of the 
non-supersymmetric attractors were studied much later \cite{gijt}. In the following we will first 
discuss the effective potential in general $N=2$ supergravity theory \cite{Ferrara:1997tw} followed
by a review of some of the known results on attractor mechanism in string theory which will be 
useful for the subsequent discussions.

The bosonic part of the Lagrangian for the $N=2$ supergravity coupled to $N$ vector multiplets
has the following form 
\cite{Ferrara:1997tw}:
\begin{eqnarray} \label{neq2sugra}
{\cal L} = 
- \frac{1}{2} R + \frac{1}{2} g_{a\bar b} \partial_\mu\phi^a\partial_\nu\bar\phi^{\bar b} h^{\mu\nu}  
- \frac{1}{4} \mu_{a b} {\cal F}^a_{\mu\nu}{\cal F}^b_{\lambda\rho}
h^{\mu\lambda}h^{\nu\rho} 
- \frac{1}{4} \nu_{ab} {\cal F}^a_{\mu\nu} *{\cal F}^b_{\lambda\rho}
h^{\mu\lambda}h^{\nu\rho} \ . \
\end{eqnarray}
Here $g_{a\bar b}$ is the moduli space metric, $h_{\mu \nu}$ is the metric on the four dimensional 
space-time, $R$ is the corresponding Ricci scalar, $X^a$ are the scalar fields and $A_{\mu}^a, a = \{0,1,\cdots n\},$ correspond to the Graviphoton and  the vector fields 
arising from the vector multiplets, ${\cal F}^a_{\mu \nu}$ are the corresponding field 
strengths. The gauge couplings $\mu_{ab}$ and $\nu_{ab}$ depend 
on the scalar fields $X^a$. For $N=2$ supergravity theories, both these matrices and the moduli 
space metric $g_{a\bar b}$ are completely determined from a prepotential.

We are interested in studying static, spherically symmetric black holes. The space-time metric for
such black holes has the form:
\begin{eqnarray}
ds^2 = e^{2 U} dt^2 - e^{-2U} \gamma_{mn} dx^mdx^n \ .
\end{eqnarray}
For such a metric, the problem of finding the black hole solution can be reduced to an effective one 
dimensional theory with a constraint. The effective potential for this one dimensional theory is given
by \cite{Ferrara:1997tw}:
\begin{eqnarray}\label{veff}
V&=&e^K\left[g^{a\bar{b}}\nabla_aW\left(\nabla_bW\right)^*+|W|^2\right]
\end{eqnarray}
where $\nabla_aW=\partial_a W + \partial_aK W$ . Here $K$ is the K\"ahler potential in the moduli
space  and $W$ is the superpotential. In terms of the $N=2$ prepotential (and the scalar fields 
$X^a$), they have the following form:
\begin{eqnarray}
K &=& - \ln{\rm Im}\left(\sum_{a=0}^N\bar{X}^{\bar a}\partial_a F(X)\right) \cr
W &=& \sum_{a=0}^N\Big(q_a X^a - p^a \partial_a F\Big) \ .
\label{kahlersuppot}
\end{eqnarray}
Here $q_a$ and $p^a$ are the electric and magnetic charges respectively. The condition for 
the existence of a regular horizon where the scalar fields $\phi^a$ take finite values is given by 
\cite{Ferrara:1997tw,gijt}:
\begin{equation}
\partial_a V = 0 \  \ {\rm at \ the \ horizon}.
\end{equation}
For the effective potential (\ref{veff}), this takes the form:
\begin{equation}\label{veffprim}
\partial_a V =
e^K \Big(g^{b\bar c} \nabla_a\nabla_b W \overline{\nabla_cW} + 2 \nabla_a W \overline{W}
+ \partial_a g^{b\bar c} \nabla_b W\overline{\nabla_c W}\Big) = 0~.
\end{equation}
This equation is trivially solved if $\nabla_aW=0$. This condition extremises the central charge
$Z = e^{K/2}W$ and corresponds to the supersymmetric attractor. However, Eq.(\ref{veffprim})
admits more general solutions which need not extremize $Z$. Such solutions correspond to 
the non-supersymmetric attractors. Unlike the supersymmetric solutions, they are generically
not stable. Their stability depends on the mass matrix $M_{ij} = (1/2)\ \partial_i\partial_j V$.
The nonsupersymmetric attractors are stable if the corresponding mass matrices $M_{ij}$ 
admit positive eigenvalues \cite{gijt}.

Let us now focus our discussion to $N=2$ supergravity theories arising from type $IIA$ string theory
compactified on a Calabi-Yau manifold. For this case, the leading order term in the prepotential is 
given by: 
 \begin{equation}
 F=D_{abc}\frac{X^aX^bX^c}{X^0} \ ,
 \end{equation} 
 where the triple intersection numbers $D_{abc}$  are completely determined by the topology of the 
 Calabi-Yau manifold:
 \begin{equation}
D_{abc}=\frac{1}{6}\int_M \alpha_a\wedge \alpha_b \wedge \alpha_c \ .
\end{equation}
Here the two forms $\alpha_a$ form a basis of the cohomology $H^2(M;{\mathbb Z})$ of the Calabi-Yau manifold $M$, over which the integration is carried out. 

Let us first consider supersymmetric attractors. They were first derived in \cite{MSW}. In this paper we will focus only on black hole configurations carrying $D0-D4$ charges.
For such configurations, the K\"ahler potential $K$ and the  superpotential  $W$ are respectively 
given by:
  \begin{eqnarray}
K&=&-\ln\left[-iD_{abc}\left(\frac{X^a}{X^0}-\frac{\bar{X}^a}{\bar{X}^0}\right)\left(\frac{X^b}{X^0}-\frac{\bar{X}^b}{\bar{X}^0}\right)\left(\frac{X^c}{X^0}-\frac{\bar{X}^c}{\bar{X}^0}\right)\right]\ ,
\end{eqnarray}   
\begin{equation}
 W=q_0X^0-3\frac{D_{ab}X^aX^b}{X^0}\ .
  \end{equation}
  Here we have introduced the matrix $D_{ab}  \equiv  D_{abc} p^c$.  For later use, we define 
  $D_a = D_{ab} p^b$ and $D = D_a p^a$. The matrix $D^{ab}$ is the inverse of $D_{ab}$. 
  The supersymmetric solution obtained by setting $\nabla_aW=0 $, is given by
\begin{equation}
   \frac{X^a}{X^0} = i p^a \sqrt{\frac{q_0}{D}} \  .
   \end{equation}
  The entropy of the black hole is given by $ S = 2\pi\sqrt{q_0D}$. 
   Clearly, this is a valid solution only when $q_0D$ is positive.
   
 The non-supersymmetric attractors for this system has been found \cite{pkt1}.  Interestingly, the 
 non-supersymmetric solution exists in the domain of the charge lattice where the supersymmetric 
 solution does not exist, i.e. when $q_0D$ is negative. They are obtained by solving the condition 
 $\partial_{a}V=0$ such that $\nabla_aW\neq 0$. For the $D0-D4$ system the solution is given by:
  \begin{equation}
   \frac{X^a}{X^0}=i p^a \sqrt{-\frac{q_0}{D}} \ . 
   \end{equation}
 Unlike the supersymmetric solutions, the non-supersymmetric ones are not guaranteed to be 
 stable. For every charge configuration, we explicitly need to check if the mass matrix is 
 positive definite. For the $D0-D4$ system the mass matrix has the following form \cite{pkt1}:
  \begin{equation}
  M = 24D{t_0}^2\left(\left(\frac{3D_a D_b}{D}-D_{ab}\right)\otimes{\bf I}+D_{ab}\otimes\sigma^3\right)
  \end{equation}
  It can be easily seen that the mass matrix  has $(n+1)$ positive and $(n-1)$ zero eigenvalues.
 One can explicitly compute terms beyond the quadratic order in the effective potential
 \cite{Nampuri:2007gv}. For the $D0-D4$ system the cubic term vanishes and the quartic 
 term is given by the difference of two positive definite terms. Thus, depending on the intersection
 numbers the quartic term can either be positive definite or negative definite. It can even give rise to 
 flat directions. In this paper we are interested to study the effect of the sub-leading terms in the 
 prepotential on the eigenvalues of the mass matrix. In the following section we will consider 
 an explicit example where the prepotential is known exactly to all orders and discuss attractor
 mechanism in this example.
  
\section{Attractors in a two-parameter model}

In the previous section, we have noticed that the leading term in the effective potential for the 
 $D0-D4$ system admits critical points with  $(n+1)$ massive and $(n-1)$ massless 
 modes. It would be interesting to see if these $(n-1)$ massless modes are lifted by sub-leading 
 corrections to the prepotential. The expression for the  prepotential has been computed 
 exactly in some one and two-parameter Calabi-Yau models by using mirror symmetry
\cite{Suranyi:1992kf,Font:1992uk,Klemm:1992tx,Candelas:1993dm,Candelas:1994hw,Govindarajan:2000vi}. 
Although 
the one-parameter models provide the simplest systems to consider, the non-supersymmetric 
attractors in this case are completely stable. For models with three or more parameters
the analysis becomes more cumbersome \cite{Hosono:1993qy}. Hence, to study the stability, 
we will focus our 
attention to the two-parameter case and study the effect of sub-leading terms in the prepotential 
on the critical points of the effective potential. 
 
 In this section we will consider one such example where the Calabi-Yau manifold is a degree 
 eight hypersurface in the weighted projective space ${\mathbb P}_4^{(1,1,2,2,2)}$. The 
 prepotential for this model has been computed exactly \cite{Candelas:1993dm} and has 
 the following form:
\begin{eqnarray} 
F&=&-\frac{1}{6X^0}\Big(8(X^1)^3+12(X^1)^2X^2\Big)-2X^1X^2-\frac{11}{3}X^0X^1-X^0X^2+\zeta(X^0)^2 \nonumber \\
&-&\frac{640}{(2\pi i)^3}{X^0}^2e^{\frac{2\pi iX^1}{X^0}}-\frac{4}{(2\pi i)^3}{X^0}^2e^{\frac{2\pi iX^2}{X^0}} + \cdots \ \ .
\label{exactprepot}
\end{eqnarray} 
Let us consider the first line in the above equation. The first term is determined by explicitly 
solving the Picard-Fuchs equation for the periods. The coefficients of all other monomials except 
the last term, are determined by requiring the symplectic invariance of the monodromy matrix 
about the boundary divisors \cite{Candelas:1993dm}. The coefficient $\zeta$ in the last term is 
not determined this way. However the effective potential does not depend on $\zeta$ and 
hence now on we will ignore this term. The second line in Eq.(\ref{exactprepot}) arises due to the 
non-perturbative contribution to the prepotential and is determined by using mirror symmetry.
 In this section we will ignore these non-perturbative contributions  and see the effect of 
 perturbative corrections. In the next section we will consider instanton contributions in a 
 more general context and  discuss their implications. 

Let us first consider the supersymmetric solution for the $D0-D4$ configuration ignoring both 
perturbative as well as non-perturbative corrections. The leading term in the prepotential is:
\begin{eqnarray}
F_0 = - \frac{1}{6X^0}\Big(8(X^1)^3+12(X^1)^2X^2\Big) \ .
\end{eqnarray}
The supersymmetric attractor is given by
\begin{eqnarray}
\label{susyleading1}
\frac{X^1}{X^0} & = & i\sqrt{\frac{-3q_0}{2\left(2p^1+3p^2\right)}}\\ \label{susyleading2}
\frac{X^2}{X^0} & = & i\frac{p^2}{p^1}\sqrt{\frac{-3q_0}{2\left(2p^1+3p^2\right)}}
\end{eqnarray}
where as the non-supersymmetric attractor is given by
\begin{eqnarray}
\frac{X^1}{X^0} & = & i\sqrt{\frac{3q_0}{2\left(2p^1+3p^2\right)}}\\
\frac{X^2}{X^0} & = & i\frac{p^2}{p^1}\sqrt{\frac{3q_0}{2\left(2p^1+3p^2\right)}}
\end{eqnarray}
It is important to note that the supersymmetric solution exists if and only if  $q_0(2 p^1 + 3 p^2) < 0$ 
where as the non-supersymmetric solution exists only when $ q_0( 2 p^1 + 3 p^2) > 0 $.

Let us now consider sub-leading corrections to the prepotential ignoring the non-perturbative part:
\begin{eqnarray}
F=-\frac{1}{6X^0}\Big(8(X^1)^3+12(X^1)^2X^2\Big)-2X^1X^2-\frac{11}{3}X^0X^1-X^0X^2+\zeta{X^0}^2 
\end{eqnarray}
We are interested in studying both supersymmetric as well as non-supersymmetric solutions 
for the $D0-D4$ system with this prepotential. Let us first consider the supersymmetric case. 
The superpotential
and the K\"ahler potential can be evaluated using the formulae (\ref{kahlersuppot}). Defining 
$x_1 = X^1/X^0$ and $x_2 = X^2/X^0$ and choosing the gauge $X^0 = 1$ we find
\begin{eqnarray}
W&=&q_ 0 + p^1 \left(4 x_ 1 (x_ 1 + x_ 2)  +2 x_ 2  +\frac{11}{3}\right) \nonumber 
 + p^ 2 (2 {x_ 1}^2 + 2 x_ 1 + 1)\\
K&=&-\ln\left(\frac{2i}{3}(x_1-\bar{x_1})^2\ \Big(2(x_1-\bar{x_1})+3(x_2-\bar{x_2})\Big)\right)
\label{wandk}
\end{eqnarray}
The condition for supersymmetry is given by  the two complex equations 
$\nabla_1 W = 0 = \nabla_2 W$. 
Using the explicit expression for $W$ and $K$ from Eq.(\ref{wandk}), and setting 
$x_1 = u_1 + i u_2 , x_2 = v_1 + i v_2$, we find, for $u_2 \neq 0$ and $ (2 u_2 + 3 v_2) \neq 0$, the real and imaginary parts of both these equations are respectively given by
\begin{eqnarray}
\label{redw1}
 u_2^2 \Big( p^2(1  + 2  u_1) + 2 p^1 (2 u_1 + v_1) \Big) 
+ 3 p^1 v_2 (1 + 2 u_1 ) (u_2 + v_2)  &=& 0\ , \\ 
\label{redw2} 
3 (p^1 v_2 - p^2 u_2) (1 + 2 u_1) + 2 p^1 u_2 (2 - 2 u_1 - 3 v_1) &=& 0\ , \\
\label{imdw1} 
 4 u_2^2 v_2 ( p^1 + p^2) + \Big( 3 q_0 + 
p^1 \{ 11 + 4 u_2^2 + 6 v_1 + 12 u_1 (u_1 + v_1)\}   \cr 
  + p^2 \{3 + 6 u_1 (1 + u_1) + 2 u_2^2\}\Big) (u_2 + v_2)  & = & 0 \ ,\\
\label{imdw2}
 3 q_0 +  p^1 \Big( 11 + 6 v_1 + 12 u_1 (u_1 + v_1) + 4 u_2 (u_2 + 3 v_2)\Big) &&  \cr
+ 3 p^2 \Big(1 + 2 u_1 (1 + u_1) - 2 u_2^2\Big)   &=& 0\ . 
\end{eqnarray}
Surprisingly the above set of equations do not admit any consistent solution for generic  values 
of the $D0-D4$ charges. This can easily be seen as follows. For $u_2\neq 0$ and 
$ 2 u_2 + 3 v_2 \neq 0$, Eqs.(\ref{imdw1}) and  (\ref{imdw2}) admit a unique solution for 
$v_1$ and $v_2$ in terms of $u_1, u_2$ and the charges $q_0,p^1,p^2$, which is given by
\begin{eqnarray}
v_1 &=& \frac{-1}{6 p^2 (1 + 2 u_1)} \Big( 
3 q_0 + 3 p^2\Big(1 + 2 u_1 (1 + u_1) + 2 u_2^2\Big) + p^1 \Big( 11 + 12 u_1^2 + 4 u_2^2\Big)\Big) \cr
v_2 &=& \frac{p^2}{p^1} u_2 \ .
\end{eqnarray}
We can substitute these expressions for $v_1$ and $v_2$ in Eqs.(\ref{redw1}) and (\ref{redw2}).
For $(1 + 2 u_1)\neq 0$ we get
\begin{eqnarray}
3 (q_0 + p^2) + 6 p^2 (u_1 + u_1^2 + u_2^2) + p^1 \Big(15 + 4 u_1 (1+u_1) + 4 u_2^2\Big) &=& 0 \ ,\\
9 (p^2)^2 (1 + 2 u_1)^2 + (p^1)^2 \Big(-11 + 12 u_1 (1+u_1) - 4 u_2^2\Big) - 3 q_0 p^1 \cr
+ p^1 p^2 \Big(9 + 42 u_1 (1 + u_1) - 6 u_2^2\Big) &=& 0\ .
\end{eqnarray}
It can easily be verified that all the solutions for the above set of equations has $u_1 = -1/2$
in contradiction to our assumption $(1 + 2 u_1)\neq 0$. We can try to solve Eqs.(\ref{redw1})-(\ref{imdw2}) eliminating different set of variables and so on. However we will reach the same 
conclusion, i.e.,  these set of equations do not admit any solution for generic values of the 
$D0-D4$ charges.

Though we do not have supersymmetric black holes in this very simple model for generic values 
of the charges, we do have consistent solutions when the black hole charges obey certain constraint.
To see this, consider the ansatz $x_1 = p^1 t$ and $x_2 = p^2 t$. For this ansatz, the susy 
conditions $\nabla_1 W = 0 = \nabla_2W$ give two complex equations involving the complex 
variable $t$:
\begin{eqnarray}
11 (p^1)^2 +14 p^1 p^2 +3 (p^2)^2 + 3 p^1 q_0 +3 p^2 q_0 +10 (p^1)^2 p^2 t 
&& \cr +9 p^1 (p^2)^2 t 
4 (p^1)^4 t^2 +10 (p^1)^3 p^2 t^2 
+6 (p^1)^2 {p^2}^2 t^2 +2(p^1)^2 p^2 \bar{t} && \cr +3 p^1 (p^2)^2 \bar{t} 
+8 (p^1)^4 |t|^2  
+20 (p^1)^3 p^2 |t|^2  +12(p^1)^2 (p^2)^2 |t|^2 &=&0 \\
11 p^1 + 3 p^2 + 3 q_0 - 4 (p^1)^2 t + 6 p^1 p^2 t +4 (p^1)^3 t^2 && \cr + 
 6 (p^1)^2 p^2 t^2 
  + 4 (p^1)^2 \bar{t} 
  + 6 p^1 p^2 \bar{t} + 8 (p^1)^3 |t|^2  + 12 (p^1)^2 p^2 |t|^2 & = & 0
\end{eqnarray}
It is straightforward to check that, both these equations coincide if and only if 
$p^2 = - 2 p^1$. In this case the equation of motion has the following simple form:
\begin{eqnarray}
5 p^1 + 3 q_0 - 8 (p^1)^2 (2 t + \bar{t}) - 8 (p^1)^3 t (t + 2 \bar{t})=0
\label{exactsusy}
\end{eqnarray}
It is now straightforward to solve the above equation for $t$. For the supersymmetric solution with 
the ansatz $x_a = p^a t$, the expression for $t$ is given by
\begin{eqnarray}
t= - \frac{1}{2p^1} - i \frac{1}{2\sqrt{2}p^1}\sqrt{11+\frac{3q_0}{p^1}} \ .  \label{susysublead}
\end{eqnarray} 
The entropy of the supersymmetric black hole can be evaluated by using the equation
\begin{equation}
\label{susyentropy}
  S = \pi e^K |W|^2 
\end{equation} and is given by
\begin{eqnarray}
S = \frac{4}{3}(p^1)^2\pi\sqrt{22+\frac{6q_0}{p^1}} \ .
\end{eqnarray}

Before we turn our attention to the non-supersymmetric attractors, it is worth emphasizing a few 
important points. First, the solution exists only if $(11 + 3 q_0/p^1) > 0$. However this is not so
different from the condition $q_0 (2 p^1 + 3 p^2) < 0$ for the existence of the supersymmetric 
solution when we 
consider only the leading term in the prepotential.  In fact both the conditions are identical in 
the limit $ |q_0| \gg |p^1|$ provided $p^2 = - 2 p^1$. As expected, the exact solution (\ref{exactsusy}) coincides with the leading order solution (\ref{susyleading2}) up to terms of order 
$(1/p^1)$. What is more surprising is the condition 
$p^2 = - 2 p^1$ imposed on the $D4$-brane charges for the existence of supersymmetric 
solution when we consider sub-leading corrections to the prepotential. Naively one would 
expect that the sub-leading terms will change the supersymmetric solution only in a small way.
However, what we observe is that the solution ceases to exist for generic values of $D4$-brane
charges. We do not know if this is an artifact of the toy example we are considering in this 
section. The curvature corrections to the $N=2$ supergravity Lagrangian (\ref{neq2sugra}) 
might restore the solutions for all values of the charges. Clearly, we need further investigation
to understand this issue.

Now we will discuss non-supersymmetric attractors in the presence of sub-leading terms in the 
prepotential. In this case we need to extremize the effective potential to find the solution. The 
effective potential for our case has been derived in \S{}A.1. We find
 \begin{eqnarray}
V&=&\Big[\Big( 6 {q_0}^2 +44 p^1 q_0 +12 p^2 q_0 +(242/3)(p^1)^2 +6(p^2)^2 + 44 p^1 p^2\Big) \cr
&+& \Big( 24 p^1 q_0 + 12 p^2 q_0 +96(p^1)^2 + 30(p^2)^2 +60 p^1 p^2 \Big) |x_1|^2 
+ 24(p^1)^2 |x_2|^2 \cr
&+& \Big( 64(p^1)^2 +24(p^2)^2 +64 p^1 p^2 \Big) |x_1|^4 
+72(p^1)^2 |x_1 x_2|^2 + \Big(12 p^1 q_0 x_2 \bar{x_1}\cr
&+&12 p^1 q_0 x_1^2+6 p^2 q_0 x_1^2+12 p^2q_0x_1
+12p^1 q_0 x_2+12 p^1 q_0 x_1 x_2+8(p^1)^2 x_2 \bar{x_1}^3 \cr
&+& 16(p^1)^2 x_1^3 \bar{x_1} 
+16 p^1 p^2 x_1^3 \bar{x_1}+24(p^2)^2 x_1^2 \bar{x_1}+40 p^1 p^2 x_1^2
   \bar{x_1}+52(p^1)^2 x_2 \bar{x_1}\cr 
&+&48 p^1 p^2 x_1 x_2
   \bar{x_1}+12(p^1)^2 x_1^2 x_2 \bar{x_2}+48(p^1)^2 x_1 x_2
   \bar{x_2}+8 p^1 p^2 x_1^3+40(p^1)^2 x_1^2 \cr &+& 9(p^2)^2
   x_1^2+38 p^1 p^2 x_1^2+12(p^2)^2 x_1+44 p^1 p^2 x_1+44
   (p^1)^2 x_2+12 p^1 p^2 x_1^2 x_2\nonumber\\
&+&32(p^1)^2 x_2 \bar{x_1}^2
+12 p^1 p^2 x_2 \bar{x_1}^2+64(p^1)^2 x_1 x_2 \bar{x_1}^2+24 p^1 p^2 x_1 x_2 \bar{x_1}^2 \cr 
&+& 24(p^1)^2 x_1^2 x_2 \bar{x_1}
+24 p^1 p^2 x_1^2 x_2 \bar{x_1}+24 p^1 p^2 x_2 \bar{x_1}+16(p^1)^2 x_1 x_2 \bar{x_1}\cr 
&+&12 p^1 p^2 x_2+36(p^1)^2 x_1
   x_2+24 p^1 p^2 x_1 x_2+C.C.\Big)\Big]\nonumber \\&&\times\frac{-i}{ (x_ 1 - \bar {x_ 1})^2 (2( x_ 1 - \bar { x_ 1}) + 3 (x_ 2 -\bar { x_ 2}))}\ .
 \end{eqnarray}
The equations of motion are found by extremizing the above potential. It is interesting to note 
that, in this case also both the equations $\partial_1V=0$ and $\partial_2V= 0$ coincide only
when the $D4$-brane charges satisfy $p^2 = - 2 p^1$. After imposing this constraint we find
\begin{multline}
30 p^1 q_0 + 9 {q_0}^2 + 32 (p^1)^6 {\bar{t}}^2 (7 t^2 + 10 |t|^2 + {\bar{t}}^2) + 
 32 (p^1)^5 \bar{t} (7 t^2 + 22 |t|^2 + 7 {\bar{t}}^2)  \\ + 
 4 (p^1)^4 (3 t^2 + 50 |t|^2 + 31 {\bar{t}}^2) + (p^1)^2 (25 - 48 q_0 (t + 2 \bar{t})) - 
 4 (p^1)^3 (3 q_0 t^2 \\ + 5 \bar{t} (8 + 3 q_0 \bar{t})   + 2 t (10 + 9 q_0 \bar{t}))=0\ .
\end{multline}
Solving this equation, we find, for the non-supersymmetric black holes
\begin{eqnarray}
t= - \frac{1}{2p^1} -\frac{i}{2\sqrt{2}p^1}\sqrt{-11-\frac{3q_0}{p^1}}\ .
\end{eqnarray}
The entropy of the non-supersymmetric black hole is found to be 
\begin{eqnarray}
S_{Nonsusy}=\frac{4}{3}(p^1)^2\pi\sqrt{-22-\frac{6q_0}{p^1}}\ .
\end{eqnarray}
The results are pretty much similar to the ones for the supersymmetric black hole attractors.  
The black 
hole solution does not exist for generic charge configuration. It exists only on the two dimensional 
sub-lattice obtained by imposing the condition $p^2 = - 2 p^1$ of the charge lattice spanned by
$q_0,p^1,p^2$.  In this sub-lattice the supersymmetric solution 
exists when $(11 + 3 p_0/p^1) > 0$ where as non-supersymmetric solution exists for 
$(11 + 3 q_0/p^1) < 0$. We need to check the stability of the non-supersymmetric black hole solution. The mass 
matrix for the black hole can be obtained by using the formula
\begin{equation}
M= \partial_{a}\partial_{\bar{ b}} V\otimes {\bf I} +\Re(\partial_{a}\partial_{b} V) \otimes {\bf \sigma^3}-    2 \Im(\partial_{a}\partial_{b}V)\otimes {\bf \sigma^1}\ .      
\end{equation}
Each of the terms are evaluated in \S{}A.1. The mass matrix is found to be 
\begin{equation}
M = \frac{\sqrt{2}(p^1)^2}{\sqrt{-11-\frac{3q_0}{p^1}}}
\left(
\begin{array}{cccc}
 12  & 0 & -6  & 0 \\
 0 & 12  & 0 & 2  \\
 -6  & 0 & 3  & 0 \\
 0 & 2  & 0 & 3 
\end{array}
\right) \ .
\end{equation}
It is straightforward to diagonalize it. The eigenvalues of the mass matrix are given by
\begin{eqnarray}
\lambda= \frac{\sqrt{2}(p^1)^2}{\sqrt{-11-\frac{3q_0}{p^1}}}
\Big(15,\frac{1}{2}\big(15+\sqrt{97}\big),
\frac{1}{2}\big(15-\sqrt{97}\big),0\Big) \ .
\end{eqnarray}
Interestingly we observe that the zero-mode still survives. Thus, the perturbative corrections 
to the prepotential restrict the allowed black hole solutions to a sub-space of the charge lattice. 
However the issue of stability for the non-supersymmetric attractors still remains to be shorted
out.

 \section{ Type \bf{IIA} at large volume with instanton corrections}

  In the previous section we  considered $D0-D4$ configuration in a two-parameter Calabi-Yau 
  model and found supersymmetric as well as non-supersymmetric attractors. In both the case, 
  the sub-leading corrections to the prepotential forced the solution to  exist only in a sub-lattice 
  of the charge lattice for the $D0-D4$ system.  When restricted to this sub-lattice, the exact 
  solution in the presence of the sub-leading terms differ from the leading solution by terms 
  of order $O(1/p^1)$. For the non-supersymmetric critical point, the 
  massless mode did not get lifted by these corrections. We would like to study the attractor 
  solution in a more general context in stead of restricting ourselves to a specific two-parameter 
  Calabi-Yau manifold. In the general case also the perturbative corrections to the prepotential 
  will impose a few constraints on the charges. Apart from the above restriction, we do not expect 
  any drastic change in the qualitative behavior of the attractor solution due to these perturbative 
  corrections. In particular, in the case of the non-supersymmetric solution, we expect a few 
  massless modes surviving these corrections. Hence, in order to lift these massless 
  directions, we would like to incorporate instanton corrections to the prepotential. In the 
  presence of the instanton term the computation for the effective potential and the mass 
  matrix is extremely tedious. Hence we will restrict ourselves to the simplest non-trivial
  case where the
  prepotential has the form
    \begin{equation}
 F=D_{abc}\frac{X^aX^bX^c}{X^0}+A(X^0)^2e^{C_aX^a/X^0}\ .
 \end{equation}
 Here the coefficients $A$ and $C_a$ in the pre-factor as well as in the exponent of the instanton
 term are constants which will depend on the details on the Calabi-Yau manifold under consideration. Throughout this 
 section we will assume that the contribution due to the exponential term is small and we will 
 ignore higher powers of the exponential term. 
 
 We will  now compute the expression for the superpotential and the K\"ahler potential for the system. For the $D0-D4$ configuration the superpotential is given by
  \begin{equation}
 W=q_0X^0-3\frac{D_{ab}X^aX^b}{X^0}-\frac{AC_ap^a}{X^0}e^{C_bX^b/X^0}\ .
  \end{equation}
  The K\"ahler potential for the system is found to have the form
\begin{eqnarray}
K&=&-\ln\Big[-iD_{abc}\left(\frac{X^a}{X^0}-\frac{\bar{X}^a}{\bar{X}^0}\right)\left(\frac{X^b}{X^0}
-\frac{\bar{X}^b}{\bar{X}^0}\right)\left(\frac{X^c}{X^0}-\frac{\bar{X}^c}{\bar{X}^0}\right)\nonumber \\
&-&i\left(\frac{X^a}{X^0}-\frac{\bar{X}^a}{\bar{X}^0}\right)\left(A\frac{\bar{X}^0}{X^0}C_a e^{C_bX^b/X^0}+\bar{A}\frac{X^0}{\bar{X}^0}\bar{C}_ae^{\bar{C}_b\bar{X}^b/ \bar{X}^0}\right)\nonumber\\
&+&2iX^0\bar{X}^0\left(Ae^{C_bX^b/X^0}-\bar{A}e^{\bar{C}_b\bar{X}^b/ \bar{X}^0}\right)
\Big]\ .
\end{eqnarray}
From now on, we will introduce $x^a = X^a/X^0$ and set the gauge $X^0 = 1$. With this choice, 
the superpotential and K\"ahler potential reads
\begin{eqnarray}
W&=&q_0-3D_{ab}x^ax^b-AC_ap^ae^{C_b x^b}\\
K&=&-\ln\Big[-i\Big(D_{abc}\left(x^a-\bar{x}^a\right)\left(x^b-\bar{x}^b\right)\left(x^c-\bar{x}^c\right)\nonumber \\
&+&\left(x^a-\bar{x}^a\right)\left(AC_a e^{C_b x^b}+\bar{A}\bar{C}_a e^{\bar{C}_b\bar{x}^b}\right)+2\left(-Ae^{C_b x^b}+\bar{A}e^{\bar{C}_b\bar{x}^b}\Big)\right)\Big]\ .
\end{eqnarray}
For convenience, we introduce the constant $T = C_a p^a$ and the function  $k(x,\bar x) =C_a\left(x^a-\bar{x}^a\right)$. We will often denote $k$ for the function $k(x,\bar x)$. In addition, we 
define 
\begin{eqnarray}
M&=&D_{abc}\left(x^a-\bar{x}^a\right)\left(x^b-\bar{x}^b\right)\left(x^c-\bar{x}^c\right)\\
L&=&A(k-2)e^{C_bx^b}-\bar{A}(\bar{k}-2)e^{\bar{C}_b\bar{x}^b}\ .
\end{eqnarray}
In terms of these quantities the superpotential and the K\"ahler potential have the simple form
\begin{eqnarray}
W&=&q_0-3D_{ab}x^ax^b-ATe^{C_b x^b}\\
K&=&-\ln\left[-i\left(M+L\right)\right]\ .
\end{eqnarray}

\subsection{Supersymmetric solution}

In this subsection we will consider the supersymmetric solution for the system. Hence we need to 
solve the equation
\begin{equation}
\nabla_aW=\partial_aW+\partial_aKW=0\ .
\end{equation}
Substituting the expression for $W$ and $K$ we get
\begin{eqnarray}
&-&\frac{1}{M+L}\Big[6MD_{a b}x^b+3M_aW^0+\Big(MTC_a-3M_aT+C_a(k-1)W^0\nonumber\\
&+&6D_{a b}x^b(k-2)\Big)Ae^{C_b x^b}+\Big(\bar{C_a}W^0-6D_{a b}x^b(\bar{k}-2)\Big)\bar{A}e^{\bar{C}_b\bar{x}^b}\Big]=0\ ,
\label{susyeqn}
\end{eqnarray}
where 
 \begin{equation}
 W^0=q_0-3D_{ab}x^ax^b \ ,
 \end{equation}
 is the leading order superpotential for the $D0-D4$ system. In the absence of the instanton 
 contribution the solution for supersymmetric attractor
 is given by 
 \begin{equation}
 x^a = i p^a \sqrt{\frac{q_0}{D}} \ . 
 \end{equation}
 We expect the solution of Eq.(\ref{susyeqn}) for $x^a$ to differ from the above by a term of order 
 $e^{C_ax^a}$. Hence we set the ansatz 
   \begin{equation}
  x^a=i\sqrt{\frac{q_0}{D}}p^a+\tilde{x}^a
 \end{equation} 
to solve Eq.(\ref{susyeqn}) and keep terms up to order $\tilde x^a$. It is convenient to consider
 $p^a\nabla_aW=0$. After simplification we get,
    \begin{eqnarray}
    6iq_0\sqrt{\frac{q_0}{D}}D_c\bar{\tilde{x^c}}+Ae^{C_b x^b}\left(3 q_0T+3 iD\sqrt{\frac{q_0}{D}}\right)
+\left(2 q_0\bar{T} + 3 iD\sqrt{\frac{q_0}{D}}\right)=0\ , \nonumber \\ 
    \end{eqnarray}
which gives
    \begin{equation}
    D_c\bar{\tilde{x}}^c=\frac{1}{6i}\sqrt{\frac{D}{q_0}}\left(\left(2\bar{T}+3i\sqrt{\frac{D}{q_0}}\right)\bar{A}e^{\bar{C}_b\bar{x}_0^b}-\left(3T+3i\sqrt{\frac{D}{q_0}}\right)Ae^{C_bx_0^b}\right)\ . 
    \end{equation}
    The entropy of the supersymmetric black hole is given by
       \begin{equation}
    S_{BH}=\pi e^K |W|^2 |_{\phi_{i0}}
    \end{equation}
    Note that 
    \begin{equation}
    |W|^2 |_{\phi_{i0}}=16q_0^2+24q_0i\sqrt{\frac{q_0}{D}}\left(D_a\bar{\tilde{x}}^a-D_a\tilde{x}^a\right)-4q_0\big(ATe^{C_b x^b_0}+\bar{A}\bar{T}e^{\bar{C}_b\bar{x}_0^b}\big)
        \end{equation}
 and
               \begin{equation}
              e^{K}~ |_{\phi_{i0}}=\frac{1} {8D{t_0}^3}+\frac{1}{16D^2{t_0}^6} \Re\big[(2Tt_0+7i)Ae^{C_b {x_0}^b}\big]\ .
              \end{equation}\\
        Substituting for  $\left(D_a\bar{\tilde{x}}^a-D_a\tilde{x}^a\right)$  in $|W|^2$, we get
        \begin{equation}
         |W|^2 |_{\phi_{i0}}=16q_0^2-\Re\big[(2T+\frac{6i}{t_0})Ae^{C_b {x_0}^b}\big]\ .
         \end{equation}
                    After simplifying a bit, we get 
              \begin{equation}
              S_{BH}=2\pi \sqrt{q_0D} - \frac{\pi \ }{{t_0}^2}{\Im }\big[Ae^{C_b {x_0}^b}\big]   \ ,           
              \end{equation}
              where $t_0=\sqrt{\frac{q_0}{D}}$ \ .

 \subsection{Effective potential with instanton corrections}
              
We will now consider the non-supersymmetric solution for the $D0-D4$ system keeping the 
instanton contribution to the prepotential. For this purpose, we need  to consider the effective 
potential 
               \begin{equation}
 V=e^K\left[g^{a\bar{b}}\nabla_aW\left(\nabla_bW\right)^*+|W|^2\right]
 \end{equation}
 and find it's critical points for which $\nabla_a W \neq 0$. In the following we will evaluate the 
 effective potential for the system and in the next subsection we will extremize it to find the 
 non-supersymmetric critical points.
 
Let us first evaluate the metric $g_{a\bar b}$  on the moduli space.
\begin{eqnarray}
g_{a\bar b} = \partial_a \partial_{\bar{b}}K&=&\frac{1}{\left(M+L\right)^2}\Big[-M\partial_{\bar{b}}\partial_aM-M\partial_{\bar{b}}\partial_aL-L\partial_{\bar{b}}\partial_aM\nonumber\\
&+&\partial_aM\partial_{\bar{b}}M+\partial_aM\partial_{\bar{b}}L+\partial_aL\partial_{\bar{b}}M\Big]\ .
\end{eqnarray}
Substituting for the expressions inside the bracket we get
\begin{eqnarray}
g_{a \bar{b}}&=&\frac{1}{\left(M+L\right)^2}\Big[6MM_{a b}-9M_aM_b+\Big(Ae^{C_b x^b}\Big(MC_aC_b+6M_{a b}(k-2)\nonumber \\
&-&3M_aC_b-3M_bC_a(k-1)\Big)+H.C\Big)\Big]\ .
\end{eqnarray}
The inverse of the metric $g_{a\bar b}$ is found to be 
\begin{eqnarray}
g^{c \bar{d}}&=&\frac{\left(M+L\right)^2}{6M^2}\Big[MM^{c d}-3\left(x^c-\bar{x}^c\right)\left(x^d-\bar{x}^d\right)-\left(\frac{1}{6M}\Big(M^2M^{q c}M^{p d}C_pC_q\right.\nonumber \\
&+&\left.3MM^{c q}C_q(x^d-{\bar{x}}^d)(2-k)+3MM^{d q}C_q(x^c-{\bar{x}}^c)(k+6)\right.\nonumber \\&+&\left.6MM^{c d}(k+2)-9k(x^c-{\bar{x}}^c)(x^d-{\bar{x}}^d)(k+6)\right.\nonumber\\&+&\left.36(x^c-{\bar{x}}^c)(x^d-{\bar{x}}^d)\Big)Ae^{C_b x^b}+H.C.\right)\Big]\ .
\end{eqnarray}
On the other hand, the covariant derivative of the superpotential is given by 
\begin{eqnarray}
\nabla_aW&=&-\frac{1}{M+L}\Big[6MD_{a b}x^b+3M_aW^0+\Big(MTC_a-3M_aT+C_a(k-1)W^0\nonumber\\
&+&6D_{a b}x^b(k-2)\Big)Ae^{C_b x^b}+\left(\bar{C_a}W^0-6D_{a b}x^b(\bar{k}-2)\right)\bar{A}e^{\bar{C}_b\bar{x}^b}\Big]\ .
\end{eqnarray}
 Substituting these terms in the expression for the effective potential and simplifying we get
     \begin{equation}
        V=Ue^K=e^K\left(U_0+U_1Ae^{C_bx^b}+ \bar{U_1} \bar{A} 
        e^{\bar{C}_b\bar{x}^b}\right) \label{veffinstanton} \ ,
             \end{equation}
             where
 \begin{eqnarray} 
  U_0&=&4|W^0|^2+6D_{a q}\left(x^a-\bar{x}^a\right)\left(x^q\bar{W}^0-\bar{x}^q W^0\right) \nonumber \\
    & +&6D_{a q}D_{b d}x^q\bar{x}^d\left(MM^{a b}-3\left(x^a-\bar{x}^a\right)\left(x^b-\bar{x}^b\right)\right) 
            \end{eqnarray}
            and 
\begin{eqnarray}
 U_1&=&\frac{1}{M}\Big(D_{a q}D_{b d}x^q\bar{x}^d\left[-3MM^{p b}C_p\left(x^a-\bar{x}^a\right)\left(k+6\right)+6MM^{a b}(k-6)\right. \nonumber \\
&+&\left.3MM^{p a}C_p\left(x^b-\bar{x}^b\right)\left(k-2\right)+9\left(k^2+2k+4\right)\left(x^a-\bar{x}^a\right)\left(x^b-\bar{x}^b\right)\right.\nonumber \\
&-&\left.M^2M^{q a}M^{p b}C_pC_q\right]+D_{b d}\bar{x}^d\left[M^2TM^{p b}C_p-3MT\left(x^a-\bar{x}^a\right)\left(k-2\right)\right.\nonumber \\&+&\left.3\left(k^2+6k-4\right))\left(x^b-\bar{x}^b\right)W^0-M^{p b}C_pW^0\left(k+10\right)\right]\nonumber \\&-&D_{a q}x^q\left[MM^{a p}C_p\bar{W}^0\left(k-2\right)+3\left(k^2+2k-4\right)\left(x^a-\bar{x}^a\right)\bar{W}^0\right]\nonumber \\&+&M\bar{W}^0T\left(k-4\right)-\left(k^2+6k-12\right)|W^0|^2\Big) \ .
\end{eqnarray}
                
\subsection{Non-supersymmetric solution}

We will now extremize the effective potential (\ref{veffinstanton}) to obtain the instanton 
corrected non-supersymmetric
solution. We set the ansatz $x^a = p^a t$ and try to solve the equation of motion with the assumption that the solution will differ from the leading order non-supersymmetric solution
$x_0^a = i p^a \sqrt{-q_0/D}$ by $O(e^{C_ax_0^a})$. Thus we assume $t = i t_0 + (m + i s)$ 
with $t_0 = \sqrt{-q_0/D}$ and keep terms up to linear order in $m$ and $s$. The equation of motion 
is a bit tedious in this case and we list some steps for the  derivation of this equation in \S{}B.1. We see that, for the above mentioned ansatz, the equation of motion takes the form:
\begin{eqnarray}
-192iD^3{t_0}^5(3m-is)+8D^2{t_0}^4\Big(3T-2it_0T^2+2{t_0}^2T^3\Big)Ae^{C_b {x_0}^b} \cr 
+8D^2{t_0}^4\Big(9\bar{T}-4it_0{\bar{T}}^2\Big)\bar{A}e^{\bar{C}_b\bar{x_0}^b} = 0\ .
\end{eqnarray}  
Solving the above equation for $m$ and $s$ we get
\begin{eqnarray}
m&=&\frac{1}{36Dt_0}{\Im}\left(T\big(-3-3i{t_0}T+T^2{t_0}^2\big)Ae^{C_b {x_0}^b}\right)
\end{eqnarray}
and
\begin{eqnarray}
s&=&-\frac{1}{12Dt_0}{\Re}\Big(T\Big(6+it_0T+{t_0}^2T^2\Big)Ae^{C_b {x_0}^b}\Big)\ .
\end{eqnarray}
For the above solution, we find 
\begin{eqnarray}
U|_{\phi_{i0}}&=&16D^2{t_0}^4-Dt_0\Re\Big(\Big(-3i+7Tt_0+{t_0}^3T^3\Big)Ae^{C_b {x_0}^b}\Big)\ ,\\
e^{K}|_{\phi_{i0}}&=&\frac{1}{8D{t_0}^3}+\frac{1}{32D^2{t_0}^6}\Re\Big(\Big(2i+8Tt_0+i{t_0}^2T^2+{t_0}^3T^3\Big)Ae^{C_b {x_0}^b}\Big)\ .
\end{eqnarray}
Hence, the entropy of the black hole is given by
\begin{eqnarray}
S=2\pi Dt_0+\frac{\pi}{2{t_0}^2}{\Re}\Big(\Big(5i+7Tt_0+i{t_0}^2T^2\Big)Ae^{C_b {x_0}^b}\Big)\ .
\end{eqnarray}

\subsection{Mass matrix}

In the preceding subsections, we have seen that it is possible to find a consistent instanton 
corrected attractor solution for both the supersymmetric as well as the non-supersymmetric 
cases. In order to know the stability of the non-supersymmetric attractor, we need to evaluate
the mass matrix 
\begin{equation}
M= \partial_{b}\partial_{\bar{ a}} V\otimes {\bf I} +\Re(\partial_{b}\partial_{a} V) \otimes {\bf \sigma^3}-2\Im(\partial_{b}\partial_{a}V)\otimes {\bf \sigma^1}      
\end{equation}
at the attractor point. The computations are really cumbersome and we give the detail 
derivation in \S{}B.2. In summary, various terms of the mass matrix are given by
\begin{eqnarray}
\partial_{\bar{b}}\partial_{a}V&=&24D{t_0}^2\Big(\frac{3D_{a }D_{b}}{D}-D_{ab}\Big)+\Big(Ae^{c_b{ x_{0}}^b }\Big(\frac{2D_{\alpha \beta}}{t_0}\Big(12i+6t_0T-2i{t_0}^2T^2\nonumber\\
&+&T^3{t_0}^3\Big)+\frac{D_{a}D_{b}}{Dt_0}\Big(-i-43t_0T+3iT^2{t_0}^2-6T^3{t_0}^3\Big)+2C_{a}C_{b}\Big(4iDt_0\nonumber\\
&-&DT{t_0}^2\Big)+2iDt_0TJ_{ab}-\frac{3D_{a}C_{b}}{2}\Big(19+2iTt_0\Big)+\frac{3D_{b}C_{a}}{2}\Big(7-8it_0T\nonumber\\&+&2T^2{t_0}^2)\Big)+H.C.\Big)\ ,
\end{eqnarray}
where $J_{ab}=D_{abs}D^{sp}C_p$. The real and imaginary parts of $(\partial_{b}\partial_{a}V)$ are found to be
\begin{eqnarray}
\Re(\partial_{b}\partial_{a}V)&=&24DD_{ab}{t_0}^2+\Big[Ae^{c_b{ x_{0}}^b }\Big(\frac{6D_{ab}}{t_0}\Big(-i+3Tt_0+2iT^2{t_0}^2\Big)\nonumber\\&+&\frac{9D_{a}D_{b}}{Dt_0}\Big(i-Tt_0-iT^2{t_0}^2\Big)+Dt_0C_{a}C_{b}\Big(-13i+7T{t_0}+iT{t_0}^2\Big)\nonumber\\
&+&\frac{3}{2}\Big(2+5it_0T-T^2{t_0}^2\Big)\Big(D_{a}C_{b}+D_{b}C_{a})\Big)+C.C.\Big]
\end{eqnarray}
and 
\begin{eqnarray}
\Im(\partial_{b}\partial_{a}V)&=&\Big[Ae^{c_b{ x_{0}}^b }\Big(\frac{4}{3}D_{ab}\Big(-3i+3T{t_0}+iT^2{t_0}^2\Big)+18iT\frac{D_{a}D_{b}}{D}\nonumber\\&+&2DJ_{ab}\Big(-i+t_0T\Big)+Dt_0C_{a}C_{b}\Big(-12-7iT{t_0}+T^2{t_0}^2\Big)\nonumber\\&+&\frac{3}{2}\Big(-3i+3Tt_0+iT^2{t_0}^2\Big)\Big(D_{a}C_{b}+D_{b}C_{a})\Big)+C.C.\Big]\ .
\end{eqnarray}

It is in general quite difficult to diagonalize the above mass matrix. In the following subsection 
we consider the special case when $n=3$ when the only non-vanishing component of $D_{abc}$ 
is given by $D_{123} = 1$. In the absence of the instanton correction, this is the STU model. In the 
next subsection we will analyze the mass matrix for this special case.

\subsection{STU model}

Consider the special case, where the number of vector multiplets is $n=3$ and the only 
non-vanishing intersection number $D_{abc}$ is $D_{123}$. For simplicity, let $D_{123} = 1$
and choose $p^1 = p^2 = p^3 = p$. In the absence of the subleading corrections, the mass matrix 
is diagonalizable and the black hole effective potential has two exactly flat directions. In the 
following subsection we will show that, both the flat directions can be lifted in the presence of
instanton term. For this purpose, consider the matrix elements of various terms appearing in the 
mass matrix:
\begin{equation}
D_{\alpha \beta}=p\left(\begin{array}{ccc}
 0 &1&1 \\
1 &  0&1 \\
1& 1  &0 \\
\end{array}\right)\nonumber\ ,
\end{equation} 
\begin{equation}
D_{\alpha}D_{ \beta}=4p^4\left(\begin{array}{ccc}
 1 &1&1 \\
1 &  1&1 \\
1& 1  &1 \\
\end{array}\right)\ ,
\end{equation} 
\begin{equation}
D_{\alpha}C_ {\beta}=2p^2\left(\begin{array}{ccc}
 C_1 &C_2&C_3 \\
C_1 &  C_2&C_3 \\
C_1& C_2  &C_3 \\
\end{array}\right)\ ,
\end{equation}
\begin{equation}
J_{\alpha \beta}=\frac{1}{2p}\left(\begin{array}{ccc}
 0 &C_1+C_2-C_3&C_1-C_2+C_3 \\
C_1+C_2-C_3+ &  0&-C_1+C_2+C_3 \\
C_1-C_2+C_3&-C_1+ C_2+C_3  &0 \\
\end{array}\right)\ .
\end{equation} 
We will scale out a factor of $(p)^2$ from the mass matrix. We will first diagonalize the mass 
matrix for the STU model using perturbative methods. Subsequently we will do the  calculation 
numerically as well as find the solution by brute force for some special choice of the parameters. 

\subsection{First order perturbation correction}

In this section we will evaluate the spectrum of the mass matrix for the STU model using first order perturbation theory. We will set the mass matrix $M$ as  $M = M_0 + M_1$ where $M_0$ corresponds to the mass matrix in the absence of instanton term and $M_1$ is the correction
piece. The matrix $M_0$ has the simple form:
\begin{eqnarray}
M_0=288t_0^2\left(
\begin{array}{cccccc}
 1 & 0 & 1 & 0 & 1 & 0 \\
 0 & 1 & 0 & 0 & 0 & 0 \\
 1 & 0 & 1 & 0 & 1 & 0 \\
 0 & 0 & 0 & 1 & 0 & 0 \\
 1 & 0 & 1 & 0 & 1 & 0 \\
 0 & 0 & 0 & 0 & 0 & 1
\end{array}
\right)\ .
\end{eqnarray}
The eigen values of the matrix are given by
\begin{equation}
288t_0^2\left\{0, 0, 1 , 1 , 1 , 3\right\}
\end{equation}
 where $t_0=\sqrt{-q_0/6p}$. The normalized eigenvectors are given by
\begin{eqnarray}
|\psi_{i}\rangle&=&\left\{-\frac{1}{\sqrt{2}},0,0,0,\frac{1}{\sqrt{2}},0\right\}, \left\{-\frac{1}{\sqrt{6}},0,\sqrt{\frac{2}{3}},0,-\frac{1}{\sqrt{6}},0\right\} ,\{0,0,0,0,0,1\}, \nonumber\\&&,\{0,0,0,1,0,0\}
\{0,1,0,0,0,0\}, \left\{\frac{1}{\sqrt{3}},0,\frac{1}{\sqrt{3}},0,\frac{1}{\sqrt{3}},0\right\} \ .
\end{eqnarray}
The expression for the matrix elements of $M_1$ are extremely lengthy for us to write it here. 
However, for some special choice of the parameters they take simple form. Note that the exponential terms in the mass matrix are $e^{C_b{x_0}^b}$ and $e^{\bar{C_b}\bar{{x_0}^b}}$, where ${x_0}^a=ip^at_0$. Let us assume that $C_1=C_2=C_3=C$ then the exponential terms take the form $e^{3it_0C}$ and $e^{-3it_0\bar{C}}$. We further assume that $C=ir$, $r$ is real. 
The correction to the eigenvalues corresponding to the zero-modes of the mass matrix $M_0$ 
can be found by diagonalizing the matrix:
   \begin{equation}
 \langle \psi_i|M_1|\psi_j\rangle=\left( \begin{array}{cc}
 m_{11}&m_{12}\\
 m_{21}&m_{22}\\
 \end{array}\right)\ .
 \end{equation}
 The eigenvalues of this matrix are given by
\begin{eqnarray}
\lambda_1&=&-\frac{36 \Im(A) e^{-3 t_0 r} (t_0 r (t_0 r (3 t_0 r+5)-5)-1)}{t_0^3}\\
\lambda_2&=&-\frac{36 \Im(A) e^{-3 t_0 r} (t_0 r (t_0 r (3 t_0 r+5)-5)-1)}{t_0^3} \ .
\end{eqnarray}
If we restrict $A=-ih$ to be negative imaginary, $h > 0$. For all $t_0r>1$  we get positive values for both $\lambda_1$ and $\lambda_2$:
\begin{eqnarray}
\lambda_1&=&\frac{36 h e^{-3 t_0 r} (t_0 r (t_0 r (3 t_0 r+5)-5)-1)}{t_0^3}\\
\lambda_2&=&\frac{36 h e^{-3 t_0 r} (t_0 r (t_0 r (3 t_0 r+5)-5)-1)}{t_0^3}
\end{eqnarray}


We can also diagonalize the full mass matrix  numerically by assuming certain values for the parameters in the mass matrix. We take the values $A=-i$,$  t_0= 1$. Fig.\ref{fig:equalc} below shows the variation of the six eigenvalues as a function of $r$. The last two plots correspond 
to the lift  of both the zero-modes.

\begin{figure}[!ht]
\centering
\includegraphics [scale=.75]{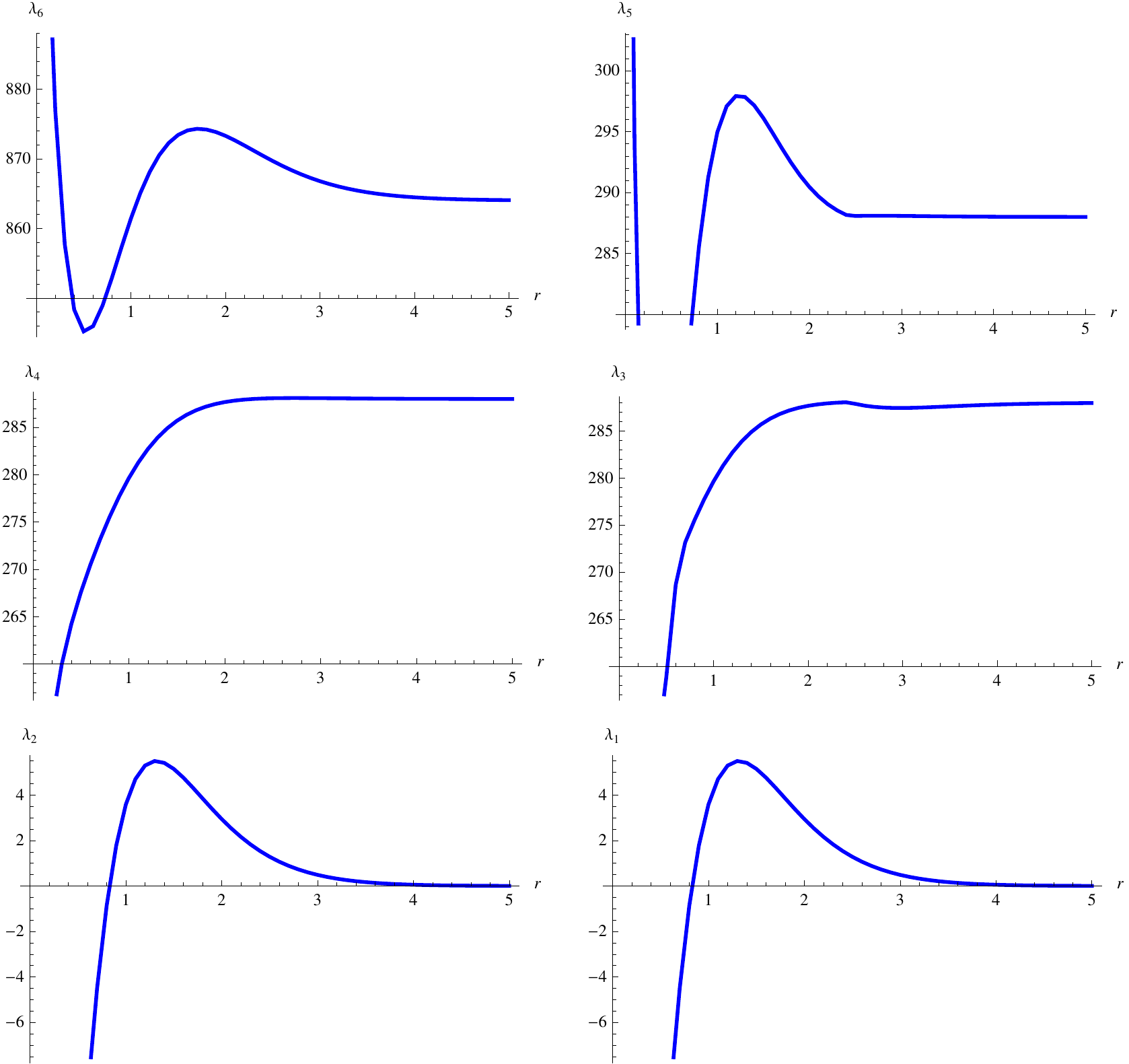}
\caption{Variation of the 6 eigenvalues with r , r varies from 0 to 5  }
\label{fig:equalc}
\end{figure} 


\subsection{Exact result}

It is in fact possible to find the eigenvalues of the mass matrix exactly for some special cases. For example, if we assume  $C_1=C_2=C_3=C$ and take $C$ to be purely imaginary, we can diagonalize the 
mass matrix for generic values of $A$. However the solution extremely lengthy and we will 
not give it here. The solution becomes much simpler if we assume that $A=ih$, where $h$ is real. The eigenvalues of the mass matrix for this case are listed below:
\begin{eqnarray}
\lambda_1&=&-36 e^{-3 t_0 r}\frac{h}{t_0}\Big(3 t_0^3 r^3+5 t_0^2 r^2-5 t_0 r-1\Big)\ ,\\
\lambda_1&=&-36 e^{-3 t_0 r}\frac{h}{t_0} \Big(3 t_0^3 r^3+5 t_0^2 r^2-5 t_0 r-1\Big)\ ,\\
\lambda_3&=&288 t_0^2-12 e^{-3 t_0 r}\frac{h}{t_0} \Big(9 t_0^3 r^3-21 t_0^2 r^2+3 t_0 r-5\Big)\ ,\\
\lambda_4&=&288 t_0^2-12e^{-3 t_0 r}\frac{h}{t_0} \Big(9 t_0^3 r^3-21 t_0^2 r^2+3 t_0 r-5\Big)\ ,\\
\lambda_5&=&288 t_0^2+4e^{-3 t_0 r}\frac{h}{t_0} \Big(81 t_0^4 r^4-189 t_0^3 r^3-99 t_0^2 r^2+192 t_0 r-20\Big)\ ,\\
\lambda_6&=&864 t_0^2-4e^{-3 t_0 r}\frac{h}{t_0} \Big(81 t_0^4 r^4-27 t_0^3 r^3+9 t_0^2 r^2-102 t_0 r+26\Big)\ .
\end{eqnarray}
The first two eigenvalues $\lambda_1$ and $\lambda_2$ correspond 
to the zero-modes in the absence of instanton terms. The remaining
four eigenvalues are corrected by exponentially suppressed terms due
to the instanton correction.
It is clear from the above expressions that by suitably adjusting the 
parameters 
the zero-modes get lifted up. 
%
\section{Conclusion} 

In this paper we have studied supersymmetric as well as non-supersymmetric attractors in
the presence of sub-leading terms in the prepotential of the $N=2$ supergravity theory. As 
a toy model, we considered the example of a two-parameter Calabi-Yau model where the 
prepotential is computed exactly to all orders using mirror symmetry. In this example, we 
considered single centered $D0-D4$ black hole solutions. Interestingly, we observed that 
in the presence of sub-leading terms, the $D0-D4$ black hole is generically destabilized. 
Stable supersymmetric
solutions exists when the $D4$-charges are restricted to the condition $p^2 = - 2 p^1$. 
The behavior of the non-supersymmetric solution is quite similar to the supersymmetric 
one. However, in this case, the massless direction still survives the perturbative corrections 
to the prepotential. In order to lift the massless mode, we considered instanton correction to 
the prepotential.  We observed that, for the case of a three parameter model, the massless
modes can in fact be lifted by the non-perturbative correction. 

Throughout the paper, we have focused our attention to single centered $D0-D4$ black hole 
configurations. The $D0-D4-D6$ black holes behave pretty much the same way as the $D0-D4$
system. Especially, for the non-supersymmetric attractors, the number of flat directions in the 
leading term of the effective black hole potential remains unchanged. It would be interesting to 
study the effect of sub-leading terms for the $D0-D4-D6$ system.  It would also be interesting 
to study the walls of marginal stability in more detail. The condition on the charges in the presence
of the sub-leading terms in the prepotential might get modified in the presence of curvature
correction to the leading order $N=2$ supergravity action. We hope to report on some of these issues in future.

\section{Acknowledgments}
We would like to thank Suresh Govindarajan and Sandip Trivedi for useful discussions. We are 
grateful to S. Lakshmi Bala for a careful reading of the manuscript. This work 
was partially supported by CEFIPRA Project No. 4104-2.


\appendix

\section{Attractors in the two-parameter model}
\subsection{Effective Potential}
In this section we give some details about the computations in section 3. 
We can calculate the K\"ahler metric using the equation $g_{a\bar{b}}=\partial_a\partial_{\bar{b}}K$ and its components are given by
\begin{eqnarray}
g_{1\bar{1}}&=&-6\frac{\left(2(x_1-\bar{x_1})^2+4(x_1-\bar{x_1})(x_2-\bar{x_2})+3(x_2-\bar{x_2})^2\right)}{(x_1-\bar{x_1})^2(2(x_1-\bar{x_1})+3(x_2-\bar{x_2}))}\\
g_{1\bar{2}}&=&-\frac{6}{(2(x_1-\bar{x_1})+3(x_2-\bar{x_2}))^2}\\
g_{2\bar{1}}&=&-\frac{6}{(2(x_1-\bar{x_1})+3(x_2-\bar{x_2}))^2}\\
g_{2\bar{2}}&=&-\frac{9}{(2(x_1-\bar{x_1})+3(x_2-\bar{x_2}))^2}
\end{eqnarray}
Inverse of $g_{a\bar{b}}$ is $g^{a\bar{b}}$ such that $g_{a\bar{b}}g^{\bar{b}c}={\delta_{a}}^c$ and the components of $g^{a\bar{b}}$ are as follows
\begin{eqnarray}
g^{1\bar{1}}&=&-\frac{(x_1-\bar{x_1})^2}{2}\\
g^{1\bar{2}}&=&\frac{(x_1-\bar{x_1})^2}{3}\\
g^{2\bar{1}}&=&\frac{(x_1-\bar{x_1})^2}{3}\\
g^{2\bar{2}}&=&-\frac{1}{3}\left(2(x_1-\bar{x_1})^2+4(x_1-\bar{x_1})(x_2-\bar{x_2})+3(x_2-\bar{x_2})^2\right)
\end{eqnarray}
The covariant derivatives of the superpotential $\nabla_iW=\partial_iW+\partial_iKW$ reads
\begin{eqnarray}
\nabla_1W&=&\Big[\Big(2p^2+4p^2x_1+4p^1(2x_1+x_2)\Big)\Big(x_1-\bar{x_1}\Big)\Big(2(x_1-\bar{x_1})+3(x_2-\bar{x_2})\Big)\nonumber\\
&-&\Big(6(p^2+q_0+2p^2x_1+2 p^2{x_1}^2)+2p^1(11+12{x_1}^2+6x_2+12x_1x_2)\Big)\nonumber\\&&\Big(x_1-\bar{x_1}+x_2-\bar{x_2}\Big)\Big]\Big[\Big(x_1-\bar{x_1}\Big)\Big(2(x_1-\bar{x_1})+3(x_2-\bar{x_2})\Big)\Big]^{-1}\\
\nabla_2W&=&\Big[\Big(p^1(2+4x_1)\Big)\Big(2(x_1-\bar{x_1})+3(x_2-\bar{x_2})\Big)-3\Big(p^2+q_0+2p^2x_1+2p^2{x_1}^2\nonumber\\&+&p^1(11/3+4{x_1}^2+2x_2+4x_1 x_2)\Big)\Big]\Big[\Big(2(x_1-\bar{x_1})+3(x_2-\bar{x_2})\Big)\Big]^{-1}
\end{eqnarray}
The effective black hole potential can be evaluated using the formula
\begin{eqnarray}
V&=&e^K\left[g^{a\bar{b}}\nabla_aW\left(\nabla_bW\right)^*+|W|^2\right]
\end{eqnarray}
After a bit tedious computations we get
\begin{eqnarray}
V&=&\Big[24 p^1 q_0 x_1 \bar{x_1}+12 p^2 q_0 x_1 \bar{x_1}+44 p^1 q_0+12 p^2
   q_0+64(p^1)^2 x_1^2 \bar{x_1}^2\nonumber \\&+&24(p^2)^2 x_1^2
   \bar{x_1}^2+64 p^1 p^2 x_1^2 \bar{x_1}^2+72(p^1)^2 x_1 x_2
   \bar{x_1} \bar{x_2}+96(p^1)^2 x_1 \bar{x_1}\nonumber\\&+&30(p^2)^2 x_1
   \bar{x_1}+60 p^1 p^2 x_1 \bar{x_1}+24(p^1)^2 x_2
   \bar{x_2}+(242/3)(p^1)^2+6(p^2)^2\nonumber \\&+&44 p^1 p^2+6 {q_0}^2+\Big(12 p^1 q_0 x_2 \bar{x_1}+12 p^1 q_0 x_1^2+6 p^2 q_0 x_1^2+12 p^2q_0x_1\nonumber\\&+&12p^1 q_0 x_2+12 p^1 q_0 x_1 x_2+8(p^1)^2 x_2
   \bar{x_1}^3+32(p^1)^2 x_2 \bar{x_1}^2+12 p^1 p^2 x_2
   \bar{x_1}^2\nonumber \\&+&64(p^1)^2 x_1 x_2 \bar{x_1}^2+24 p^1 p^2 x_1
   x_2 \bar{x_1}^2+16(p^1)^2 x_1^3 \bar{x_1}+16 p^1 p^2 x_1^3
   \bar{x_1}\nonumber \\&+&24(p^2)^2 x_1^2 \bar{x_1}+40 p^1 p^2 x_1^2
   \bar{x_1}+52(p^1)^2 x_2 \bar{x_1}+24(p^1)^2 x_1^2 x_2
   \bar{x_1}\nonumber\\&+&24 p^1 p^2 x_1^2 x_2 \bar{x_1}+24 p^1 p^2 x_2
   \bar{x_1}+16(p^1)^2 x_1 x_2 \bar{x_1}+48 p^1 p^2 x_1 x_2
   \bar{x_1}\nonumber\\&+&12(p^1)^2 x_1^2 x_2 \bar{x_2}+48(p^1)^2 x_1 x_2
   \bar{x_2}+8 p^1 p^2 x_1^3+40(p^1)^2 x_1^2+9(p^2)^2
   x_1^2\nonumber\\&+&38 p^1 p^2 x_1^2+12(p^2)^2 x_1+44 p^1 p^2 x_1+44
   (p^1)^2 x_2+12 p^1 p^2 x_1^2 x_2\nonumber\\&+&12 p^1 p^2 x_2+36(p^1)^2 x_1
   x_2+24 p^1 p^2 x_1 x_2+C.C.\Big)\Big]\nonumber \\&&\times\frac{-i}{ (x_ 1 - \bar {x_ 1})^2 (2( x_ 1 - \bar { x_ 1}) + 3 (x_ 2 -\bar { x_ 2}))}\ .
 \end{eqnarray}

The non-supersymmetric solution is given by the solution of the the set of equations\\
\begin{equation}
\partial_iV_{eff}=0
\end{equation}
We have two complex equations $\partial_1V_{eff}=0$ and $\partial_2V_{eff}=0$. After substituting the ansatz $x_1=p^1t$ and $x_2=p^2t$ these equations read,
\begin{eqnarray}
&&18 p^2 (p^2 + q_0)^2 + 16 (p^1)^7 {\bar{t}}^2 (7 t^2 + 10 |t|^2  + {\bar{t}}^2) + 
 64 (p^1)^6 p^2 \bar{t}^2 (7 t^2 + 10 |t|^2  + {\bar{t}}^2) \nonumber\\&+& 
 6 p^1 (p^2 + q_0) \Big(25 p^2 + 3 q_0 + 3 (p^2)^2 (3 t + 5 \bar{t})\Big) + 
 2 (p^1)^4 \Big(18 (p^2)^3 {\bar{t}}^2 (7 t^2 + 10 |t|^2 \nonumber \\&+& {\bar{t}}^2)+
    6 q_0 (t^2 + 6 |t|^2  + 5 {\bar{t}}^2) + 
    12 (p^2)^2 \bar{t} (13 t^2 + 38 |t|^2 + 9 {\bar{t}}^2) + 
    p^2 (55 t^2 + 378 |t|^2\nonumber\\&+ &299 {\bar{t}}^2)\Big) + 
 4 (p^1)^5 \Big({\bar{t}}^2 (54 + 20 p^2 \bar{t} + 21 (p^2)^2 {\bar{t}}^2) + 
    2 |t|^2 (34 + 46 p^2 \bar{t} + 105 (p^2)^2 {\bar{t}}^2) \nonumber\\&+& 
    t^2 (10 + 32 p^2 \bar{t} + 147 (p^2)^2 {\bar{t}}^2)\Big) - 
 2 (p^1)^2 \Big[66 q_0 + 18 (p^2)^3 (t^2 + 8 |t|^2 + 5 {\bar{t}}^2) \nonumber\\&+& 
    p^2 \Big(187 + 6 q_0 (5 t + 7 \bar{t})\Big) + 
    3 (p^2)^2 \Big(3 q_0 t^2 + 3 \bar{t} (23 + 5 q_0 \bar{t}) + t (43 + 18 q_0 \bar{t})\Big)\Big]\nonumber\\& +& 
 (p^1)^3 \Big[242 + 36 (p^2)^3 \bar{t} (5 t^2 + 15 |t|^2 + 4 {\bar{t}}^2) + 
    3 (p^2)^2 (37 t^2 + 262 |t|^2 + 181 {\bar{t}}^2)\nonumber\\& +& 
    2 p^2 \Big(15 q_0 t^2 + 10 t (11 + 9 q_0 \bar{t}) + \bar{t} (154 + 75 q_0 \bar{t})\Big)\Big]=0
\end{eqnarray}
\begin{eqnarray}
&&18(p^2 + q_0)^2 + 16 (p^1)^6 {\bar{t}}^2 (7 t^2 + 10 |t|^2+ {\bar{t}}^2) + 
 12 p^1 (p^2 + q_0) \Big(11 + 3 p^2 (t + 3 \bar{t})\Big) \nonumber \\&+& 
 16 (p^1)^5 \bar{t} \Big({\bar{t}}^2 (4 + 3 p^2 \bar{t}) + 2 |t|^2  (-1 + 15 p^2 \bar{t}) + 
    t^2 (-2 + 21 p^2 \bar{t})\Big) \nonumber \\&+& 
 4(p^1)^4 \Big(\bar{t}^2 (56 + 60 p^2 \bar{t} + 9 (p^2)^2 {\bar{t}}^2) + 
    3 t^2 (4 + 4 p^2 \bar{t} + 21 (p^2)^2 {\bar{t}}^2) \nonumber \\&+& 
    2 |t|^2(32 + 36 p^2 \bar{t} + 45 (p^2)^2 {\bar{t}}^2)\Big) + 
 2 (p^1)^3 \Big(t^2 (33 p^2 + 6 q_0 + 72 (p^2)^2 \bar{t}) \nonumber \\&+& 
    \bar{t} (44 + 237 p^2 \bar{t} + 30 q_0 \bar{t} + 108 (p^2)^2 {\bar{t}}^2) + 
    2 t (-22 + 99 p^2 \bar{t} + 18 q_0 \bar{t} + 126 (p^2)^2 \bar{t}^2)\Big)\nonumber\\& +&
 (p^1)^2 \Big[242 - 24 q_0 t + 24 q_0 \bar{t} + 9 (p^2)^2 (3 t^2 + 26 |t|^2 + 27 {\bar{t}}^2) + 
    6 p^2 \Big(3 q_0 t^2 \nonumber\\&+& 18 t (1 + q_0 \bar{t}) + 5 \bar{t} (14 + 3 q_0 \bar{t})\Big)\Big]=0
\end{eqnarray}
Substituting the condition $p^2=-2p^1$ both the equations reduce to the following equation,
\begin{multline}
30 p^1 q_0 + 9 {q_0}^2 + 32 (p^1)^6 {\bar{t}}^2 (7 t^2 + 10 |t|^2 + {\bar{t}}^2) + 
 32 (p^1)^5 \bar{t} (7 t^2 + 22 |t|^2 + 7 {\bar{t}}^2)  \\ + 
 4 (p^1)^4 (3 t^2 + 50 |t|^2 + 31 {\bar{t}}^2) + (p^1)^2 (25 - 48 q_0 (t + 2 \bar{t})) - 
 4 (p^1)^3 (3 q_0 t^2 \\ + 5 \bar{t} (8 + 3 q_0 \bar{t})   + 2 t (10 + 9 q_0 \bar{t}))=0\ .
\end{multline}
Solving this equation, we find, for the non-supersymmetric black holes
\begin{eqnarray}
t= - \frac{1}{2p^1} -\frac{i}{2\sqrt{2}p^1}\sqrt{-11-\frac{3q_0}{p^1}}\ .
\end{eqnarray}
\subsection{Mass matrix}
For the stability analysis of the attractor we need to calculate the mass matrix
\begin{equation}
M= \partial_{b}\partial_{\bar{ a}} V\otimes {\bf I} +\Re(\partial_{b}\partial_{a} V) \otimes {\bf \sigma^3}-2\Im(\partial_{b}\partial_{a}V)\otimes {\bf \sigma^1}\ .    
\end{equation}
For the mass matrix the second derivatives should be evaluated at the attractor point and we know that the solution exist only if we impose the condition $p^2=-2p^1$. We have imposed this condition, evaluated the second derivatives and thereafter we have substituted the ansatz $x_1=p^1t$ and $x_2=p^2t$. Here we list the various second derivative terms.
\begin{eqnarray}
\partial_1\partial_1V&=&\frac{3i}{4 (p^1)^5 (t - \bar{t})^5}\Big (30 p^1 q_0 + 9 {q_0}^2 +
   32 (p^1)^6 {\bar{t}}^2 (7 t^2 + 10 t \bar{t} + {\bar{t}}^2)\nonumber\\& +& 
   32 (p^1)^5 \bar{t} (7 t^2 + 22 t \bar{t} + 7 {\bar{t}}^2) + 
   4 (p^1)^4 (3 t^2 + 50 t \bar{t} + 31 {\bar{t}}^2) \nonumber\\&+& (p^1)^2 (25 - 48 q_0 (t + 2 \bar{t})) - 
   4 (p^1)^3 \Big(3 q_0 t^2 + 5 \bar{t} (8 + 3 q_0 \bar{t}) \nonumber\\&+& 2 t (10 + 9 q_0 \bar{t})\Big)\Big)
\end{eqnarray}
\begin{eqnarray}
\partial_1\partial_2V&=&\frac{i}{8 (p^1)^5 (t - \bar{t})^5}\Big(-30 p^1 q_0 - 9 {q_0}^2 + (p^1)^2 (-25 + 144 q_0 \bar{t})\nonumber\\&+&
   12 (p^1)^4 (t^2 - 2 t \bar{t} - 27 {\bar{t}}^2) - 
   32 (p^1)^6 {\bar{t}}^2 (-7 t^2 + 14 t \bar{t} + 11 {\bar{t}}^2) \nonumber\\&-& 
   32 (p1)^5 \bar{t} (-7 t^2 + 14 t \bar{t} + 29 {\bar{t}}^2) + 
   12 (p^1)^3 \Big(20 \bar{t} + q_0 (-t^2 \nonumber\\&+& 2 t \bar{t} + 11 {\bar{t}}^2)\Big)\Big)
\end{eqnarray}
\begin{eqnarray}
\partial_2\partial_2V&=&\frac{3i}{16 (p^1)^5 (t - \bar{t})^5)} \Big(30 p^1 q_0 + 9 {q_0}^2 + 
   32 (p^1)^6 {\bar{t}}^2 (7 t^2 + 10 t \bar{t} + {\bar{t}}^2)\nonumber\\&+& 
   32 (p^1)^5 \bar{t} (7 t^2 + 22 t\bar{t} + 7 {\bar{t}}^2) + 
   4 (p^1)^4 (3 t^2 + 50 t \bar{t} + 31 {\bar{t}}^2)\nonumber\\&+& (p^1)^2 (25 - 48 q_0(t + 2 \bar{t})) - 
   4 (p^1)^3 \Big(3 q_0 t^2 + 5 \bar{t} (8 + 3 q_0 \bar{t})\nonumber\\&+& 2 t (10 + 9 q_0 \bar{t})\Big)\Big)
\end{eqnarray}
\begin{eqnarray}
\partial_1\partial_{\bar{1}}V&=&\frac{-3i}{4 (p^1)^5 (t - \bar{t})^5)}
 \Big(30 p^1 q_0 + 9 {q_0}^2+96(p^1)^6 t\bar{t}(t^2 + 4 t \bar{t} + {\bar{t}}^2)\nonumber\\& + &
    12 (p^1)^4 (3 t^2 + 22 t \bar{t} + 3 {\bar{t}}^2) + 
    48 (p^1)^5 (t^3 + 11 t^2 \bar{t} + 11 t{\bar{t}}^2 + {\bar{t}}^3) \nonumber\\&+& 
    (p^1)^2 (25 - 72 q_0 (t + \bar{t})) - 12 (p^1)^3 (t + \bar{t}) (10 + 3 q_0 (t + \bar{t}))\Big)
\end{eqnarray}
\begin{eqnarray}
\partial_1\partial_{\bar{2}}V&=&\frac{i}{8 (p^1)^5 (t - \bar{t})^5)} \Big(30 p^1 q_0 + 9 {q_0}^2 + 
   28 (p^1)^4 (5 t^2 + 2 t \bar{t} + 5 {\bar{t}}^2) \nonumber\\&+& 
   32 (p^1)^6 t \bar{t} (5 t^2 + 8 t \bar{t} + 5 {\bar{t}}^2) + 
   16 (p^1)^5 (5 t^3 + 31 t^2 \bar{t} + 31 t {\bar{t}}^2 + 5 {\bar{t}}^3)\nonumber\\& + &
   (p^1)^2 (25 - 72 q_0 (t + \bar{t})) - 
   12 (p^1)^3 \Big(q_0 t^2 + 10 t (1 + q_0 \bar{t}) \nonumber\\&+& \bar{t} (10 + q_0 \bar{t})\Big)\Big)
\end{eqnarray}
\begin{eqnarray}
\partial_2\partial_{\bar{2}}V&=&\frac{-3i}{16 (p^1)^5 (t - \bar{t})^5)
 } \Big(30 p^1 q_0 + 9 {q_0}^2 + 96 (p^1)^6 t \bar{t} (t^2 + 4 t \bar{t} + {\bar{t}}^2)\nonumber \\& +& 
    12 (p^1)^4 (3 t^2 + 22 t \bar{t} + 3 {\bar{t}}^2) + 
    48 (p^1)^5 (t^3 + 11 t^2 \bar{t} + 11 t {\bar{t}}^2 + {\bar{t}}^3) \nonumber\\&+& 
    (p^1)^2 (25 - 72 q_0 (t + \bar{t})) - 12 (p^1)^3 (t + \bar{t}) (10 + 3 q_0 (t + \bar{t}))\Big)
\end{eqnarray}
Upon substituting the solution $t= - \frac{1}{2p^1} -\frac{i}{2\sqrt{2}p^1}\sqrt{-11-\frac{3q_0}{p^1}}$ these terms take the form
\begin{eqnarray}
\partial_1\partial_{1}V&=&0
\end{eqnarray}
\begin{eqnarray}
\partial_1\partial_{2}V&=&\frac{-4\sqrt{2}(p^1)^2}{\sqrt{-11-\frac{3q_0}{p^1}}}
\end{eqnarray}
\begin{eqnarray}
\partial_2\partial_{2}V&=&0
\end{eqnarray}
\begin{eqnarray}
\partial_1\partial_{\bar{1}}V&=&\frac{12\sqrt{2}(p^1)^2}{\sqrt{-11-\frac{3q_0}{p^1}}}
\end{eqnarray}
\begin{eqnarray}
\partial_1\partial_{\bar{2}}V&=&\frac{-2\sqrt{2}(p^1)^2}{\sqrt{-11-\frac{3q_0}{p^1}}}
\end{eqnarray}
\begin{eqnarray}
\partial_2\partial_{\bar{2}}V&=&\frac{3\sqrt{2}(p^1)^2}{\sqrt{-11-\frac{3q_0}{p^1}}}
\end{eqnarray}
Substituting this terms in the expression for mass matrix we get
\begin{equation}
M = \frac{\sqrt{2}(p^1)^2}{\sqrt{-11-\frac{3q_0}{p^1}}}
\left(
\begin{array}{cccc}
 12  & 0 & -6  & 0 \\
 0 & 12  & 0 & 2  \\
 -6 & 0 & 3  & 0 \\
 0 & 2  & 0 & 3 
\end{array}
\right)
\end{equation}

\section{Instanton corrections}

\subsection{Non-supersymmetric solution}

  In this section we will work out the non-supersymmetric solution by brute force method. First we will obtain the non-supersymmetric solution by imposing $x^a=p^at$ and assuming  $ t=m+i\left(t_0+s\right) $ then we will solve for $m$ and $s $ from the equation              
 \begin{equation}
 \partial_{a}V= e^K\left(\partial_{a}U+\partial_{a}KU\right)=0
 \end{equation}
We have
\begin{eqnarray}
\partial_{a}U&=&\partial_{a}U_0+\left(C_{a}U_1+\partial_{a}U_1\right)Ae^{C_b x^b}+\partial_{a}\bar{U_1}\bar{A}e^{\bar{C}_b\bar{x}^b}\\
\partial_{a}K&=&\frac{-1}{M+L}\Big[3M_{a}+C_{a}(k-1)Ae^{C_b x^b}+\bar{C_{a}}\bar{A}e^{\bar{C}_b\bar{x}^b}\Big]
\end{eqnarray}
Substituting these expressions in $\partial_aV$ we get
\begin{eqnarray}
e^{-K}\partial_{a}V&=&\frac{1}{M+L}\Big[M\partial_{a}U_0-3M_{a}U_0+L\partial_{a}U_0\nonumber\\&+&\Big(MC_{a}U_1+M\partial_{a}U_1
-C_{a}(k-1)U_0-3M_{a}U_1\Big)Ae^{C_b x^b}\nonumber\\&+&\Big(M\partial_{a}U_2-\bar{C_{a}}U_0-3M_{a}U_2\Big)\bar{A}e^{\bar{C}_b\bar{x}^b}\Big]
\end{eqnarray}
It is convenient to solve $p^a\partial_aV=0$, so we contract $\partial_aV$ with $p^a$ and substitute for L we get
\begin{eqnarray}
e^{-K}p^a\partial_{a}V&=&\frac{p^a}{M+L}\Big[M\partial_{a}U_0-3M_{a}U_0+\Big(\partial_{a}U_0(k-2)+MC_{a}U_1+M\partial_{a}U_1\nonumber \\
&-&C_{a}(k-1)U_0-3M_{a}U_1\Big)Ae^{C_b x^b}+\Big(-\partial_{a}U_0(\bar{k}-2)+M\partial_{a}\bar{U_1}\nonumber\\&-&\bar{C_{a}}U_0
-3M_{a}\bar{U_1}\Big)\bar{A}e^{\bar{C}_b\bar{x}^b}\Big]
\end{eqnarray}
We have calculated all the terms in $e^{-K}p^a\partial_{a}V$ and they are as follows,
\begin{eqnarray}
p^{a}(M\partial_{a}U_0-3M_{a}U_0)&=&-192iD^3{t_0}^5\left(3m-is)\right)\\
p^{a}\partial_{a}U_0(k-2)&=&-48iD^2{t_0}^3(iTt_0-1)\\
p^{a}MC_{a}U_1&=&-16iD^2T{t_0}^4(-3i-Tt_0+iT^2{t_0}^2)\\
p^{a}M\partial_{a}U_1&=&-8iD^2{t_0}^3(4iTt_o+3T^2{t_0}^2-3)\\
-p^{a}C_{a}(k-1)U_0&=&-16D^2T{t_0}^4(2iTt_0-1)\\
-3p^{a}M_{a}U_1&=&24D^2{t_0}^3(-3i-Tt_0+iT^2{t_0}^2)\\
-p^{a}\partial_{a}U_0(\bar{k}-2)&=&-48iD^2{t_0}^3(i\bar{T}t_0+1)\\
p^{a}M\partial_{a}\bar{U_1}&=&-8iD^2{t_0}^3(8i\bar{T}t_0+{\bar{T}}^2{t_0}^2+3)\\
-p^{a}\bar{C_{a}}U_0&=&-16D^2\bar{T}{t_0}^4\\
-3p^{a}M_{a}\bar{U_1}&=&24D^2{t_0}^3(3i-\bar{T}t_0-i{\bar{T}}^2{t_0}^2)
\end{eqnarray}
Where $t_0=\sqrt{\frac{-q_0}{D}}$. Substituting the above terms we get
\begin{eqnarray}
e^{-K}p^a\partial_{a}V&=&\frac{1}{M+L}\Big[-192iD^3{t_0}^5(3m-is)+8D^2{t_0}^4\Big(3T-2it_0T^2\nonumber\\&+2&{t_0}^2T^3\Big)Ae^{C_b {x_0}^b}
+8D^2{t_0}^4\Big(9\bar{T}-4it_0{\bar{T}}^2\Big)\bar{A}e^{\bar{C}_b\bar{x_0}^b}\Big]
\end{eqnarray}
$\partial_{\alpha}V= 0$ implies that
\begin{eqnarray}
3m-is&=&\frac{1}{24iDt_0}\Big[\Big(3T-2it_0T^2+2{t_0}^2T^3\Big)Ae^{C_b {x_0}^b}+\Big(9\bar{T}-4it_0{\bar{T}}^2\Big)\bar{A}e^{\bar{C}_b\bar{x_0}^b}\Big]\nonumber\\
\end{eqnarray} 

Solving for $m$ and $s$ we get
\begin{eqnarray}
m&=&\frac{1}{36Dt_0}{\Im}\left(T\big(-3-3i{t_0}T+T^2{t_0}^2\big)Ae^{C_b {x_0}^b}\right)
\end{eqnarray}
and
\begin{eqnarray}
s&=&-\frac{1}{12Dt_0}{\Re}\left(T\big(6+it_0T+{t_0}^2T^2\big)Ae^{C_b {x_0}^b}\right)\ .
\end{eqnarray}
For the above solution, we find 
\begin{eqnarray}
U|_{\phi_{i0}}&=&16D^2{t_0}^4-Dt_0\Re\big((-3i+7Tt_0+{t_0}^3T^3)Ae^{C_b {x_0}^b}\big)\ ,\\
e^{K}|_{\phi_{i0}}&=&\frac{1}{8D{t_0}^3}+\frac{1}{32D^2{t_0}^6}\Re\big((2i+8Tt_0+i{t_0}^2T^2+{t_0}^3T^3)Ae^{C_b {x_0}^b}\big)\ .
\end{eqnarray}
Hence, the entropy of the black hole is given by
\begin{eqnarray}
S=2\pi Dt_0+\frac{\pi}{2{t_0}^2}{\Re}\Big(\big(5i+7Tt_0+i{t_0}^2T^2\big)Ae^{C_b {x_0}^b}\Big)\ .
\end{eqnarray}

\subsection{Mass matrix }

  The matrix elements are given by the second derivative of the effective potential. The required terms are $e^{-K_0}\partial_{b}\partial_a V$ and  $e^{-K_0}\partial_{\bar{b}}\partial_a V$. 
\begin{equation}
\partial_{b}\partial_{a}V=\Big( \partial_{b}\partial_{a}U+ \partial_{b}\partial_{a}KU+\partial_{a}K\partial_{b}U\Big)e^K\ .
\end{equation}
Using the expression for $U$ we can write
\begin{eqnarray}
\partial_{b}\partial_{a}U&=&\partial_{b}\partial_{a}U_0+Ae^{c_b{ x_{0}}^b}\Big(\partial_{b}\partial_{a}U_1+C_{a}C_{b}U_1+C_{b}\partial_{a}U_1+C_{a}\partial_{b}U_1\Big)\nonumber\\&+&\bar{A}e^{\bar{c}_b\bar{x}_{0}^b}\Big( \partial_{b}\partial_{a}\bar{U_1}\Big)
 \end{eqnarray}

Using the equation of motion we can rewrite
\begin{eqnarray}
 \partial_{b}\partial_{a}K+\partial_{a}K\partial_{b}&=&\partial_{b}\partial_{a}K- \partial_{a}K\partial_{b}K
\end{eqnarray}
After doing little bit algebra we can conclude,
   \begin{eqnarray}
  \partial_{b}\partial_{a}K- \partial_{a}K\partial_{b}K&=&\frac{-1}{M+L}\Big(6M_{a b}+kC_{a}C_{b}Ae^{C_b {x}^b}\Big)\\
&=&H_0+H_1Ae^{C_b {x}^b}
\end{eqnarray}
Using this expression we can write
\begin{eqnarray}
(\partial_{b}\partial_{a}K- \partial_{a}K\partial_{b}K)U&=&H_0U_0+(H_1U_0+H_0U_1)Ae^{C_b {x_0}^b}+H_0\bar{U_1}\bar{A}e^{\bar{C}_b\bar{x}^b}\nonumber\\
 \end{eqnarray}
Using the above equations we can write
 \begin{eqnarray}
  \partial_{b}\partial_{a}Ve^{-K}&=&\partial_{b}\partial_{a}U_0+H_0U_0+Ae^{c_b{ x_{0}}^b}\Big(\partial_{b}\partial_{a}U_1+C_{a}C_{b}U_1\nonumber \\
 &+&C_{b}\partial_{a}U_1+C_{a}\partial_{b}U_1+H_0U_1+H_1U_0\Big)\nonumber \\
 &+&\bar{A}e^{\bar{c}_b\bar{x}_{0}^b} \Big(\partial_{b}\partial_{a}U_2+H_0\bar{U_1}\Big)
 \end{eqnarray}
                   
We have evaluated each term in the above expression at the critical value and their values are listed below,
\begin{eqnarray}
\partial_{b}\partial_{a}U_0&=& D_{a b} \Big[T\Big(10+5iTt_0-\frac{1}{3}T^2{t_0}^2\Big)Ae^{c_b{ x_{0}}^b}\nonumber \\
&+&\bar{T}\Big(2+3i\bar{T}t_0+\frac{7}{3}\bar{T}^2{t_0}^2\Big)\bar{A}e^{\bar{c}_b\bar{x}_{0}^b}\Big]\\
H_0U_0&=&24D{t_0}^2D_{a b}+\frac{2D_{a b}}{t_0}\Re\Big[\Big(6i-i{t_0}^2T^2-{t_0}^3T^3\Big)Ae^{c_b{ x_{0}}^b}\Big]\\
\partial_{b}\partial_{a}U_1&=&\frac{3D_{a b}}{t_0}\Big(-i+5t_0T+3i{t_0}^2T^2\Big)+\frac{9D_{a}D_{b}}{Dt_0}\Big(i-3Tt_0-i{t_0}^2T^2\Big)\nonumber\\
&+&2J_{a b}\Big(D+iDTt_0\Big)+\frac{3}{2}\Big(D_{a}C_{b}+D_{b}C_{a}\Big)\Big(7+2iTt_0\Big)\nonumber\\
&-&9iDt_0C_{a}C_{b}\\
C_{a}C_{b}U_1&=&2D{t_0}C_{a}C_{b}\Big(-3i-Tt_0+iT^2{t_0}^2\Big) \\
 C_{a}\partial_{b}U_1&=&Dt_0\Big(-5i+6T{t_0}\Big)C_{a}C_{b} +3\Big(-1+3iTt_0-T^2{t_0}^2\Big)D_{b}C_{a} \\
C_{b}\partial_{a}U_1&=&Dt_0\Big(-5i+6T{t_0}\Big)C_{a}C_{b} +3\Big(-1+3iTt_0-T^2{t_0}^2\Big) D_{a}C_{b}\\
H_0U_1&=&(-3i+t_0T+i{t_0}^2T^2)\frac{3D_{a b}}{t_0}\\
H_1U_0&=&4TD{t_0}^2C_{a}C_{b}
\end{eqnarray}
\begin{eqnarray}
\partial_{a}\partial_{b}\bar{U_1}&=&3\Big(i+5\bar{T}t_0-3i\bar{T}^2{t_0}^2\Big)\frac{D_{a b}}{t_0}+\frac{9D_{a}D_{b}}{Dt_0}\Big(-i+\bar{T}t_0+i\bar{T}^2{t_0}^2\Big)\nonumber \\
&+&2D\Big(-1+i\bar{T}t_0\Big)\bar{J}_{a b}-\frac{3}{2}\Big(1+2it_0\bar{T}\Big)\Big(D_{a}\bar{C}_{b}+D_{b}\bar{C}_{a}\Big)\nonumber\\
&+&iDt_0\bar{C}_{a}\bar{C}_{b}\\
H_0U_2&=&(3i-\bar{T}t_0-i\bar{T}^2{t_0}^2)\frac{3D_{a b}}{t_0}
\end{eqnarray}
    Adding up the above results we get
 \begin{eqnarray}
e^{-K}\partial_{b}\partial_{a}V&=&24DD_{a b}{t_0}^2+\Big[\frac{2D_{a b}}{t_0} \Big(-3i+11t_0T+8iT^2{t_0}^2-\frac{2}{3}T^3{t_0}^3\Big)\nonumber \\
&+&\frac{9D_{a}D_{b}}{Dt_0}\Big(i-3t_0T-iT^2{t_0}^2\Big)+C_{a}C_{b}Dt_0\Big(-25i+14T{t_0}\nonumber\\&+&2iT^2{t_0}^2\Big)
+2D\Big(1+it_0T\Big)J_{a b}+\frac{3}{2}\Big(5-2T^2{t_0}^2+8iTt_0\Big)\nonumber\\&&\Big(D_{a}C_{b}+D_{b}C_{a}\Big)\Big] Ae^{c_b{ x_{0}}^b}
+ \Big[2\Big(3i+7\bar{T}t_0-4i\bar{T}^2{t_0}^2\nonumber\\&+&\frac{2}{3}\bar{T}^3{t_0}^3\Big)\frac{D_{a b}}{t_0}+\frac{9D_{a}D_{b}}{Dt_0}\Big(-i+\bar{T}t_0+i\bar{T}^2{t_0}^2\Big)\nonumber \\
 &+&2D\Big(-1+i\bar{T}t_0\Big)\bar{J_{a b} }+iD t_0\bar{C}_{a} \bar{C}_{b}-\frac{3}{2}\Big(1+i\bar{T}t_0\Big)\nonumber \\&&\Big(D_{a}\bar{C}_{b}+D_{b}\bar{C}_{a}\Big)\Big] \bar{A}e^{\bar{c}_b\bar{x}_{0}^b}                  
 \end{eqnarray}
Consider the term  $\partial_{\bar{b}}\partial_{a}V$ in the mass matrix. Using the equations of motion we can write it as
\begin{equation}
\partial_{\bar{b}}\partial_{a}V=e^K\Big(\partial_{\bar{b}}\partial_{a}KU+\partial_{\bar{b}}\partial_{a} U-(\partial_{a}K   )(\partial_{\bar{b}}K)U\Big) 
 \end{equation}
Using the expression for $U$ we can write,
 \begin{eqnarray}
 \partial_{\bar{b}}\partial_{a} U&=&\partial_{\bar{b}}\partial_{a} U_0+ Ae^{c_b{ x_{0}}^b} \left(\partial_{\bar{b}}\partial_{a} U_1+C_{a}\partial_{\bar{b}}U_1\right)+\bar{A}e^{\bar{C}_b\bar{x}_{0}^b} \left(\partial_{\bar{b}}\partial_{a} \bar{U_1}+\bar{C}_{b}\partial_{a}\bar{U_1}\right)\nonumber \\ 
\end{eqnarray}
After doing little bit algebra it can be shown that
\begin{eqnarray}     
\partial_{\bar{b}}\partial_{a}K-\partial_{a}K\partial_{\bar{b}}K&=&\frac{1}{M+L}\Big(6M_{a b}+C_{a}C_{b}Ae^{C_bx^b}\nonumber\\&-&\bar{C_{a}}\bar{C_{b}}\bar{A}e^{\bar{C_b}\bar{x^b}}\Big)\nonumber\\
&=&G_0+G_1Ae^{C_bx^b}+G_2\bar{A}e^{\bar{C_b}\bar{x^b}}\\
\end{eqnarray}
We have introduced $G_0$, $G_1$ and $G_2$ for notational simplicity. Multiplying with $U$ gives
\begin{eqnarray}
(\partial_{\bar{b}}\partial_{a}K-\partial_{a}K\partial_{\bar{b}}K)U&=&G_0U_0+(G_1U_0+G_0U_1)Ae^{C_bx^b}+(G_2U_0+G_0U_2)\bar{A}e^{\bar{C_b}\bar{x^b}}\nonumber\\
\end{eqnarray}
Substituting for each terms in $\partial_{\bar{b}}\partial_{a}V$ we get
\begin{eqnarray}
e^{-K}  \partial_{\bar{b}}\partial_{a}V&  =&\partial_{\bar{b}}\partial_{a}U_0+G_0U_0+Ae^{c_b{ x_{0}}^b}\Big(G_1U_0+ G_0U_1+\partial_{\bar{b}}\partial_{a}U_1+C_{a}\partial_{\bar{b}}U_1\Big)\nonumber\\
&+&\bar{A}e^{\bar{c}_b\bar{x_0}_{0}^b}\Big(G_0\bar{U_1}+ G_2U_0+\partial_{\bar{b}}\partial_{a}\bar{U_1}+\bar{C}_{b} \partial_{a}\bar{U_1}\Big)\nonumber\\                  
\end{eqnarray}  
We have evaluated each term in the above expression explicitly at the critical value and they are listed below.
\begin{eqnarray}
  U_0G_0&=&-24D{t_0}^2D_{ab}+\frac{2D_{ab}}{t_0}\Re\Big[\Big(-6i+i{t_0}^2T^2+{t_0}^3T^3\Big)Ae^{C_bx^b}\Big]\\
\partial_{\bar{b}}\partial_{a}U_0&=&72{t_0}^2D_{a}D_{b}\nonumber\\&+&2T\Big(D_{ab}-\frac{6D_aD_b}{D}\Big)\Re\Big[\Big(6+iTt_0+T^2{t_0}^2\Big)e^{C_b{x_0}^b}\Big]
\end{eqnarray}
\begin{eqnarray}
G_1U_0&=&2iDt_0C_{a}C_{b}\\
G_0U_1&=&\frac{-3D_{ab}}{t_0}\Big(-3i-Tt_0+iT^2{t_0}^2\Big)\\
\partial_{\bar{b}}\partial_{a}U_1&=&3\Big(7i+Tt_0-iT^2{t_0}^2\Big)\frac{D_{ab}}{t_0}+\Big(-i-7Tt_0+9iT^2{t_0}^2\Big)\frac{D_{a}D_{b}}{Dt_0}\nonumber \\&+&2iDt_0TJ_{ab}+5iDt_0C_{a}C_{b}-\frac{3}{2}\Big(19+2iTt_0\Big)D_{a}C_{b}\nonumber\\&+&\frac{3}{2}\Big(5-2iTt_0\Big)D_{b}C_{a}\\
C_{a}\partial_{\bar{b}}U_1&=&Dt_0\Big(i-2T{t_0}\Big)C_{a}C_{b}+3\Big(1-3iTt_0+T^2{t_0}^2\Big)D_{b}C_{a}\\
U_0G_2&=&-2iDt_0\bar{C}_{a}\bar{C}_{a}\\
G_0\bar{U_1}&=&\frac{-3D_{ab}}{t_0}\Big(3i-\bar{T}t_0-i{\bar{T}}^2{t_0}^2\Big)\\
\partial_{\bar{c}}\partial_{a}\bar{U_1}&=&\frac{3D_{ab}}{t_0}\Big(-7i+\bar{T}t_0+i\bar{T}^2{t_0}^2\Big)+\frac{D_{a}D_{b}}{Dt_0}\Big(i-7\bar{T}t_0-9i\bar{T}^2{t_0}^2\Big)\nonumber \\&-&2iD\bar{T}t_0\bar{J}_{ab}-5iDt_0\bar{C}_{a}\bar{C}_{b}+\frac{3}{2}\Big(5+2i\bar{T}t_0\Big)D_{a}\bar{C}_{b}\nonumber\\&+&\frac{3}{2}\Big(-19+2i\bar{T}t_0)D_{b}\bar{C}_{a}\\
\bar{C}_{b}\partial_{a}U_2&=&-Dt_0\Big(i+2\bar{T}{t_0}\Big)\bar{C}_{a}\bar{C}_{b}+3\Big(1+3it_0\bar{T}+\bar{T}^2{t_0}^2\Big)D_{a}\bar{C}_{b}
\end{eqnarray}

Adding up the terms we get
\begin{eqnarray}
 e^{-K}\partial_{\bar{b}}\partial_{a}V&=&24D{t_0}^2\Big(\frac{3D_{a}D_{b}}{D}-D_{ab}\Big)\nonumber\\&+&\Big[Ae^{c_b{ x_{0}}^b }\Big(\frac{2D_{ab}}{t_0}\Big(12i+6t_0T-2i{t_0}^2T^2+T^3{t_0}^3\Big)\nonumber\\
&+&\frac{D_{a}D_{b}}{Dt_0}\Big(-i-43t_0T+3iT^2{t_0}^2-6T^3{t_0}^3\Big)\nonumber\\&+&2Dt_0C_{a}C_{b}\Big(4i-2T{t_0}\Big)+2iDt_0TJ_{a b} -\frac{3}{2}D_{a}C_{b}\Big(19+2iTt_0\Big)\nonumber\\&+&\frac{3}{2}D_{b}C_{a}\Big(7-8it_0T+2T^2{t_0}^2\Big)\Big)+H.C.\Big]
  \end{eqnarray}
 The real and imaginary part of $\Re(\partial_{b}\partial_{a}V)$, $\Im(\partial_{b}\partial_{a}V)$ are given by
\begin{eqnarray}
\Re(\partial_{b}\partial_{a}V)&=&24DD_{ab}{t_0}^2+\Big[Ae^{c_b{ x_{0}}^b }\Big(\frac{6D_{ab}}{t_0}(-i+3Tt_0+2iT^2{t_0}^2\Big)\nonumber\\&+&\frac{9D_{a}D_{b}}{Dt_0}\Big(i-Tt_0-iT^2{t_0}^2\Big)\nonumber\\&+&Dt_0\Big(-13i+7T{t_0}+iT^2{t_0}^2\Big)C_{a}C_{b}\nonumber\\&+&\frac{3}{2}\Big(2+5it_0T-T^2{t_0}^2\Big)\Big(D_{a}C_{b}+D_{b}C_{a})\Big)+C.C.\Big]\\
\Im(\partial_{b}\partial_{a}V)&=&\Big[Ae^{c_b{ x_{0}}^b }\Big(\frac{4TD_{ab}}{t_0}\Big(-i+T{t_0}+\frac{1}{3}iT^2{t_0}^2\Big)\nonumber\\&+&18it_0T\frac{D_{a}D_{b}}{Dt_0}+2D\Big(-i+t_0T\Big)J_{ab}\nonumber\\&+&Dt_0\Big(-12-7iT{t_0}+T^2{t_0}\Big)C_{a}C_{b}\nonumber\\&+&\frac{3}{2}\Big(-3i+iT^2{t_0}^2+3Tt_0\Big)\Big(D_{a}C_{b}+D_{b}C_{a}\Big)\Big)+C.C.\Big]
\end{eqnarray}

 \end{document}